\documentclass[acmsmall]{acmart}
\renewcommand\footnotetextcopyrightpermission[1]{} % Remove doi and price for arxiv

\setcopyright{none}
\usepackage{booktabs}   %% For formal tables:
\usepackage{subcaption} %% For complex figures with subfigures/subcaptions
\usepackage{array}
\newcolumntype{$}{>{\global\let\currentrowstyle\relax}}
\newcolumntype{^}{>{\currentrowstyle}}

\usepackage[ligature, inference, shorthand]{semantic}
\usepackage{amsmath}
\usepackage{xcolor}
\usepackage{alltt}
\usepackage{xspace}
\colorlet{punct}{red!60!black}
\definecolor{background}{HTML}{EEEEEE}
\definecolor{delim}{RGB}{20,105,176}
\colorlet{numb}{magenta!60!black}

\usepackage{listings}
\lstdefinelanguage{json}{
	basicstyle=%
	  \normalfont
	  \ttfamily
	  \footnotesize\lst@ifdisplaystyle\scriptsize\fi,
	numberstyle=\scriptsize,
	stepnumber=1,
	numbersep=8pt,
  morestring=[b]',
  morestring=[b]",
  stringstyle=\color{black},
	showstringspaces=false,
	breaklines=true,
    escapeinside={(*}{*)},
	literate=
	*{0}{{{\color{numb}0}}}{1}
	{1}{{{\color{numb}1}}}{1}
	{2}{{{\color{numb}2}}}{1}
	{3}{{{\color{numb}3}}}{1}
	{4}{{{\color{numb}4}}}{1}
	{5}{{{\color{numb}5}}}{1}
	{6}{{{\color{numb}6}}}{1}
	{7}{{{\color{numb}7}}}{1}
	{8}{{{\color{numb}8}}}{1}
	{9}{{{\color{numb}9}}}{1}
	{:}{{{\color{punct}{:}}}}{1}
	{,}{{{\color{punct}{,}}}}{1}
	{\{}{{{\color{delim}{\{}}}}{1}
	{\}}{{{\color{delim}{\}}}}}{1}
	{[}{{{\color{delim}{[}}}}{1}
	{]}{{{\color{delim}{]}}}}{1}
	{true}{{{\color{numb}true}}}{4}
	{false}{{{\color{numb}false}}}{5}
	{null}{{{\color{numb}null}}}{4},
}

\newcommand*{\delim}[1]{\texttt{\color{delim}{#1}}}
\newcommand*{\punct}[1]{\texttt{\color{punct}{#1}}}

\definecolor{mpcolor}{rgb}{0.1,0.9,0.1}

\definecolor{ahcolor}{rgb}{0.30,0.65,1.00}

\definecolor{mhcolor}{rgb}{1.00,0.00,0.00}

\mathlig{|->}{\mapsto}
\mathlig{<=}{\leq}
\mathlig{>=}{\geq}

\DeclareMathOperator{\valid}{JValid}
\DeclareMathOperator{\sub}{JSubSchema}
\DeclareMathOperator{\eq}{JEquivSchema}

\DeclareMathOperator{\jtypes}{\mathit{Jtypes}}
\DeclareMathOperator{\jprimitive}{\mathit{Jprimitive}}
\DeclareMathOperator{\jstruct}{\mathit{Jstructure}}

\DeclareMathOperator{\inhabited}{\mathit{inhabited}}
\DeclareMathOperator{\subrange}{\mathit{subNumber}}
\DeclareMathOperator{\alldisjointitems}{\mathit{allDisjointItems}}

\DeclareMathOperator{\kw}{\mathit{kw}}
\DeclareMathOperator{\dom}{\mathit{dom}}
\DeclareMathOperator{\default}{\mathit{default}}
\DeclareMathOperator{\typeof}{\mathit{typeOf}}

\newcommand{\canonical}{\mathrm{canonical}}

\newcommand{\typestyle}[1]{\text{\lstinline|#1|}}

\newcommand{\jstr}{\text{\ttfamily{string}}}
\newcommand{\jminlen}{\text{\ttfamily{minLength}}}
\newcommand{\jmaxlen}{\text{\ttfamily{maxLength}}}
\newcommand{\jpattern}{\text{\ttfamily{pattern}}}

\newcommand{\jint}{\text{\ttfamily{integer}}}
\newcommand{\jnum}{\text{\ttfamily{number}}}
\newcommand{\jmin}{\text{\ttfamily{minimum}}}
\newcommand{\jmax}{\text{\ttfamily{maximum}}}

\newcommand{\jmulof}{\text{\ttfamily{multipleOf}}}

\newcommand{\jarray}{\text{\ttfamily{array}}}
\newcommand{\jitems}{\text{\ttfamily{items}}}
\newcommand{\jadditems}{\text{\ttfamily{additionalItems}}}
\newcommand{\jminitems}{\text{\ttfamily{minItems}}}
\newcommand{\jmaxitems}{\text{\ttfamily{maxItems}}}
\newcommand{\juitems}{\texttt{uniqueItems}}

\newcommand{\jobj}{\text{\ttfamily{object}}}
\newcommand{\jprops}{\text{\ttfamily{properties}}}
\newcommand{\jaddp}{\text{\ttfamily{additionalProperties}}}
\newcommand{\jpatp}{\text{\ttfamily{patternProperties}}}
\newcommand{\jminp}{\text{\ttfamily{minProperties}}}
\newcommand{\jmaxp}{\text{\ttfamily{maxProperties}}}
\newcommand{\jreq}{\text{\ttfamily{required}}}
\newcommand{\jdeps}{\text{\ttfamily{dependencies}}}

\newcommand{\jbool}{\text{\ttfamily{boolean}}}

\newcommand{\jfalse}{\text{\ttfamily{false}}}
\newcommand{\jnull}{\text{\ttfamily{null}}}

\newcommand{\type}{\text{\ttfamily{type}}}
\newcommand{\enum}{\text{\ttfamily{enum}}}
\newcommand{\anyof}{\text{\ttfamily{anyOf}}}
\newcommand{\allof}{\text{\ttfamily{allOf}}}
\newcommand{\oneof}{\text{\ttfamily{oneOf}}}
\newcommand{\jnot}{\text{\ttfamily{not}}}
\newcommand{\jref}{\text{\ttfamily{\$ref}}}

\newcommand{\wpt}{\emph{WP}}
\newcommand{\kub}{\emph{K8s}}
\newcommand{\lale}{\emph{Lale}}

\newcommand{\subschema}{\emph{jsonsubschema}}
\newcommand{\issubset}{\emph{issubset}}

\begin{document}

\pagestyle{plain} % remove vol, article, pub. date for arxiv

%% Title information 
\title[Type Safety with JSON Subschema]{Type Safety with JSON Subschema}         %% [Short Title] is optional;
                                        %% when present, will be used in
                                        %% header instead of Full Title.
% \titlenote{with title note}             %% \titlenote is optional;
                                        %% can be repeated if necessary;
                                        %% contents suppressed with 'anonymous'
% \subtitle{Subtitle}                     %% \subtitle is optional
% \subtitlenote{with subtitle note}       %% \subtitlenote is optional;
                                        %% can be repeated if necessary;
                                        %% contents suppressed with 'anonymous'

%% Author information
%% Contents and number of authors suppressed with 'anonymous'.
%% Each author should be introduced by \author, followed by
%% \authornote (optional), \orcid (optional), \affiliation, and
%% \email.
%% An author may have multiple affiliations and/or emails; repeat the
%% appropriate command.
%% Many elements are not rendered, but should be provided for metadata
%% extraction tools.

\author[A. Habib]{Andrew Habib}
\affiliation{\institution{TU Darmstadt, Germany}}
\email{andrew.a.habib@gmail.com}

\author[A. Shinnar]{Avraham Shinnar}
\affiliation{\institution{IBM Research, USA}}
\email{shinnar@us.ibm.com}

\author[M. Hirzel]{Martin Hirzel}
\affiliation{\institution{IBM Research, USA}}
\email{hirzel@us.ibm.com}

%% \author[A. Shinnar]{Avraham Shinnar}
%% \affiliation{\institution{IBM Research, USA}}
%% \email{shinnar@us.ibm.com}

%% \author[M. Hirzel]{Martin Hirzel}
%% \affiliation{\institution{IBM Research, USA}}
%% \email{hirzel@us.ibm.com}

\author[M. Pradel]{Michael Pradel}
\affiliation{\institution{University of Stuttgart, Germany}}
\email{michael@binaervarianz.de}

\begin{abstract}
JSON is a popular data format used pervasively in web APIs, cloud
computing, NoSQL databases, and increasingly also machine learning.
JSON Schema is a language for declaring the structure of valid JSON
data. There are validators that can decide whether a specific JSON document is
valid with respect to a schema.
%Unfortunately, like all instance-based
%testing, these validators can only show the presence and never the
%absence of a bug.
However, validators cannot show that all documents valid under one schema are also valid under another schema.
This paper presents a technique to address this limitation:
JSON subschema checking, which can be used for static type checking with
JSON Schema.
Deciding whether one schema is a subschema of another is non-trivial because of the richness of the JSON Schema specification language.
Given a pair of schemas, our approach
first canonicalizes and simplifies both schemas, then reasons about the subschema question on the canonical forms, dispatching simpler subschema queries to
type-specific checkers. We apply an implementation of our subschema checking algorithm to
8,548 pairs of real-world JSON schemas from different domains, demonstrating
that it can decide the subschema question for most schema pairs and is
always correct for schema pairs that it can decide. We hope that our
work will bring more static guarantees to hard-to-debug domains, such as
cloud computing and artificial intelligence.
\end{abstract}

%% 2012 ACM Computing Classification System (CSS) concepts
%% Generate at 'http://dl.acm.org/ccs/ccs.cfm'.

%% End of generated code

%% Keywords
%% comma separated list
% \keywords{keyword1, keyword2, keyword3}  %% \keywords are mandatory in final camera-ready submission

%% \maketitle
%% Note: \maketitle command must come after title commands, author
%% commands, abstract environment, Computing Classification System
%% environment and commands, and keywords command.
\maketitle

\thispagestyle{plain} % remove vol, article, pub. date for arxiv

\section{Introduction}\label{sec:intro}

% What is JSON?
\emph{JSON} (JavaScript Object Notation) is a data serialization
format that is widely adopted to store data on disk or send it
over the network. 
%Derived from JavaScript, JSON is both human- and machine-readable, 
%and there are now JSON parsers for many programming languages.
The format supports primitive data types, such as strings, numbers, 
and Booleans, and two possibly nested data structures: arrays, which represent lists of values, and objects, which represent maps of key-value pairs.
%The data types can be nested, e.g., to have an array of two objects that each map a key to some primitive value.
%
% Applications of JSON
JSON is used in numerous applications.
It is the most popular data exchange format in web APIs, ahead of
XML~\cite{rodriguez_et_al_2016}.
Cloud-hosted applications also use JSON pervasively, e.g., in micro-services
that communicate via JSON data~\cite{newman_2015}.
On the data storage side, not only do traditional database management
systems, such as Oracle, IBM DB2, MySQL, and PostgreSQL, now support JSON,
but two of the most widely deployed NoSQL database 
management systems, MongoDB and Cloudant, are entirely based on JSON.
Beyond web, cloud, and database applications, JSON is also gaining
adoption in artificial intelligence (AI)~\cite{hirzel_et_al_2019,smith_et_al_2019}.

% Schemas and schema validation
With the broad adoption of JSON as a data serialization format
soon emerged the need for a way to describe how a JSON document should look.
For example, a web API that consumes JSON data can avoid unexpected behavior if it knows
the structure of the data it receives.
To describe a JSON document, \emph{JSON Schema} allows users to declaratively 
define the structure of nested values (documents) via types (schemas)~\cite{pezoa_et_al_2016}.
We adopt the value space definition for \emph{type} as being
a set of possible values~\cite{parnas_shore_weiss_1976}.
A JSON Schema \emph{validator} checks whether a JSON document $d$
conforms to a schema~$s$, denoted~\mbox{$d:s$}.
There are libraries with JSON Schema validators for many programming
languages, and they are widely used to make software more reliable.

% Limitations of schema validation and how they affect applications
Since these $d:s$ validation checks happen at runtime, they can usually
only detect problems late during deployment and production.
In cloud applications, a wrongly-structured document can cause hard-to-debug
failures, it may go unnoticed and simply cause undesired behavior, or even
worse, it may exploit a security vulnerability.
In AI, a machine learning \emph{pipeline} is a
graph of operators for preprocessing and
prediction~\cite{buitinck_et_al_2013}.
Any mismatch between training data and a pipeline, 
between production data and training data, or between data in adjacent 
steps of a pipeline can cause crashes or poor predictive
performance~\cite{Breck2019}.
Sometimes, dynamic data validity checks of the form $d:s$ only
trigger after earlier 
pipeline steps have already performed costly computations.

% Subschemas to the rescue
Given how essential dynamic $d:s$ checks are, we argue that JSON
schemas could be even more useful if they were also checked statically.
The quintessential static property is whether one schema is a \emph{subschema} of another.
We say that a schema $s$ is a subschema of a schema $t$,
denoted~\mbox{$s<:t$}, if all
documents that validate against $s$ also validate against~$t$.
Subschema checks of the form $s<:t$ can find the same mistakes as dynamic 
schema validation, but with less wasted compute and human time, as they can statically rule out entire classes of errors.

% Practical relevance of subschema checking
JSON subschema checking has various practical applications, such as statically reasoning about breaking changes of web APIs and type-checking machine learning pipelines before running them.
As one example, consider two independently evolving services that communicate over a RESTful API, where both services, as well as the schema that describes the data they exchange, are evolving~\cite{DBLP:conf/icws/LiXLZ13,DBLP:journals/jss/EspinhaZG15}.
JSON subschema checking could detect a number of likely errors before deployment by checking whether the schemas of the services remain compatible. 
As another example, consider a machine learning pipeline where the data, as well as the input and output of operations, are specified as JSON schemas~\cite{hirzel_et_al_2019,smith_et_al_2019}.
JSON subschema checking could type-check the entire pipeline before running it, saving precious developer time caused by bugs that otherwise may be detected only after hours of computation.

% Why subschema checking is difficult
JSON schemas have several features that make subschema checking
difficult. Schemas for primitive types involve non-trivial features,
such as regular expressions for strings and \jmulof{} for numbers.
Schemas for compound types involve \juitems{} constraints for arrays
and regular expressions for object property names. Furthermore, JSON
schemas can be composed using schema conjunction, disjunction, and
negation, which requires handling of complex, compound types. Finally, the
\enum{} type is entangled with other types, further exacerbating the
above-mentioned difficulties.
As a result of these and other features of JSON Schema, a simple lexical check or a structural comparison of JSON schemas is insufficient to address the subschema question.

\begin{figure}
	\centering
	\includegraphics[scale=.65]{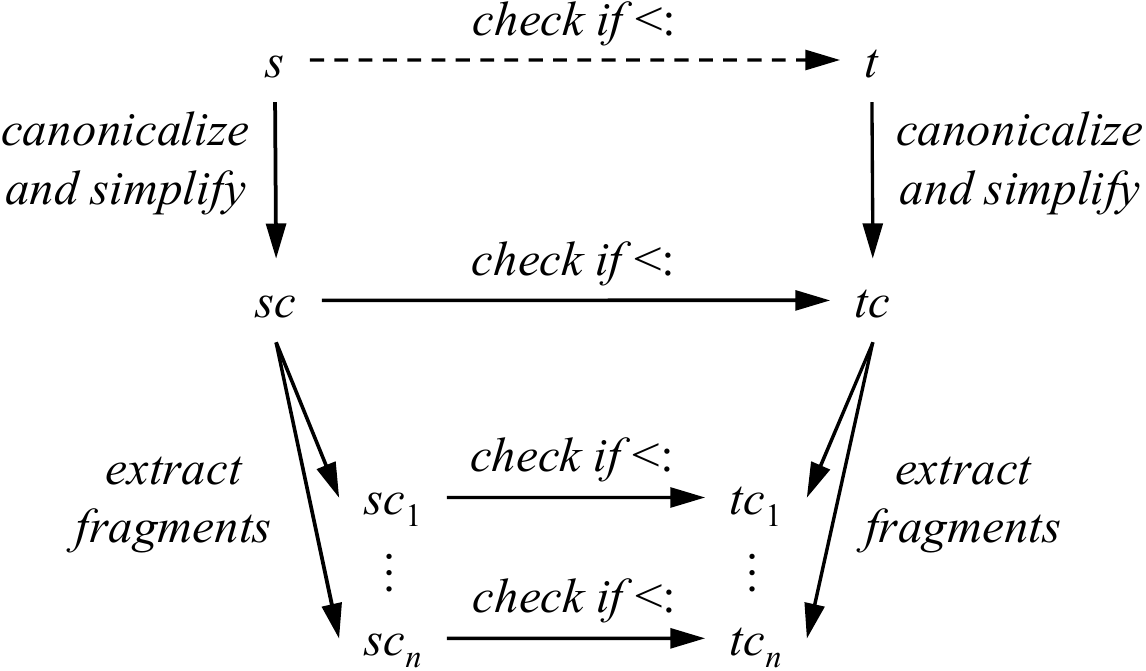}
    \vspace*{-2mm}
	\caption{\label{fig:galois}Overview of JSON subschema checker.}
\end{figure}

% This paper in a nutshell
This paper presents \subschema{}, a formally described and fully implemented subschema checker
for JSON Schema.
Figure~\ref{fig:galois} gives a high-level overview of our algorithm.
To check whether $s<:t$, the checker first canonicalizes and simplifies
$s$ to $\textit{sc}$ and $t$ to $\textit{tc}$.  This preprocessing
disentangles different cases to consider during subschema checking, while
preserving the schema semantics.  Next, our algorithm extracts
corresponding fragments $\textit{sc}_i$ and~$\textit{tc}_i$.
These fragments are type-homogeneous, i.e., they refer to only one basic
JSON type each. This homogeneity makes it possible to use separate rules to
check whether $\textit{sc}_i<:\textit{tc}_i$ for each type.

% Claims; properties of the approach
Our algorithm is sound, i.e., whenever it gives an answer about the subtype relation of two given schemas, then this answer is correct.
In principle, the algorithm is incomplete, i.e., for some pairs of subschemas, \subschema{} refuses to answer the question whether one schema is a subtype of the other.
However, the approach handles various challenging cases, including almost all that appear in practice.
Our evaluation with 8,548 pairs of real-world JSON schemas shows that the algorithm handles the vast majority of all pairs (94\%) and that all the given answers are correct.

% Prior work
The most closely related prior work is an open-source project called
\textsf{is-json-schema-subset} developed concurrently with our work,
but it only handles a small fraction of the features of JSON
Schema~\cite{haggholm_2019}. 
The problem of subschema checking has also been explored for XML, where it
is called schema containment~\cite{tozawa_hagiya_2003}. However, that
approach treats XML schemas as tree automata, which has been shown to
be insufficient for JSON schemas because JSON Schema is more expressive than tree automata~\cite{pezoa_et_al_2016}.

% Contributions
In summary, this paper makes the following contributions:

\begin{itemize}
  \item A canonicalizer and simplifier that converts a schema $s$
    into a schema $\textit{sc}$ that is simpler to check yet permits the
    same set of documents (Sections \ref{sec:canonicalization}
    and~\ref{sec:simplification}).

  \item A subschema checker for canonicalized JSON schemas that uses
    separate subschema checking rules for each basic JSON type
    (Section~\ref{sec:subtyping}).

  \item Empirical evidence on 8,548 pairs of JSON schemas taken from
    real-world applications in the web, cloud computing, and
    AI. We show that the implementation of our
    algorithm successfully answers the subschema question in most
    cases, and when it does, yields correct answers in reasonable time
    (Section~\ref{sec:evaluation}).
\end{itemize}

Our JSON subschema checker is available as open-source (link elided
for double-blind review). We also have plans for an artifact
submission.

%Overall, we hope that our work helps bring the blessings of static type safety
%to applications based on JSON and JSON Schema, e.g., in cloud computing and AI systems.

\section{Background}
\label{sec:bg}

This section briefly describes JSON data and JSON Schema validation.

\subsection{JSON Data}
\label{sub:json}

JSON was conceived as a light-weight, text-based, and
programming-language agnostic data interchange
format~\cite{jsonECMA}.

\begin{definition}[$\jprimitive$]
        The primitive JSON types are:
        $$\jprimitive = \{\jbool, \jnull, \jnum, \jint, \jstr\}$$
\end{definition}

Besides primitive types, JSON has two structured types:
ordered lists of values (arrays) and unordered maps of key-value pairs
(objects).

\begin{definition}[$\jstruct$]
        The structured JSON types are:
        $\jstruct = \{\jarray, \jobj\}$
\end{definition}

\begin{definition}[$\jtypes$]\label{def:jtypes}
        The set of JSON data types is:
        $\jtypes = \jprimitive \cup \jstruct$
\end{definition}

A \emph{valid JSON document} (or value) is either a value of one of the basic
types, or an array whose elements are valid JSON documents, or an
object mapping string keys to valid JSON documents.
%% Any value that has a type $\tau \in \jtypes$ is a valid JSON document. 
Some examples of valid JSON documents include
\lstinline|true|, \lstinline|null|, \lstinline|5|, \lstinline|'ab'|,
\lstinline|[]|, \lstinline|[1, 2, 3]|, \lstinline|[.5, {}, 'a']|,
and \lstinline|{'foo': 1, 'bar': [true, '']}|.

Finally, JSON documents can use JSON references to point to data from
the same or other JSON documents.  For example, the reference
\lstinline|{'$ref':'#'}| points to the root of the current document and
the reference \lstinline|{'$ref':'#/p/q}'| assumes that the root is an object
whose \lstinline|'p'| property is a nested object and points to its
\lstinline|'q'| property.

\subsection{JSON Schema}
\label{sub:jsonSch}

\begin{table*}\small
  \def\arraystretch{0.95}
  \caption{\label{table:jschema} Overview of JSON Schema (version draft04).}
  \vspace*{-3mm}
  { \footnotesize
  \texttt{null} and \lstinline|boolean| are basic types without keywords.
  \lstinline|$ref| refers to another schema, and \lstinline|$ref:{'#'}| points to the root of the  meta-schema, thus indicating a nested sub-schema.
  \lstinline|default:{}| indicates the top schema, which validates for all values.
  }
  \\[2mm]
  \begin{tabular}{@{}p{13em}p{7.5em}p{21em}@{}}
    \toprule
	\textbf{Description} & \textbf{Keyword} & \textbf{Meta-schema}\\
    \midrule
    \multicolumn{3}{@{}l@{}}{Schemas for type \lstinline|string|, e.g., \lstinline|\{type:'string', minLength: 1, pattern: '[a-z]*'\}|:} \\
    \midrule
    Restrict length of the string & \lstinline|minLength| & \lstinline|{type: 'integer', minimum: 0, default: 0}| \\
    & \lstinline|maxLength| & \lstinline|{type: 'integer', minimum: 0}| \\
    String must match a pattern & \lstinline|pattern| & \lstinline|{type: 'string', format: 'regex'}| \\
    \midrule
    \multicolumn{3}{@{}l@{}}{Schemas for types \lstinline|number| and \lstinline|integer|, e.g., \lstinline|\{type:'number', minimum: 0.0, multipleOf: 0.1\}|:} \\
    \midrule
    Restrict range of values  & \lstinline|minimum|          & \lstinline|{type: 'number'}|\\
    & \lstinline|maximum|          & \lstinline|{type: 'number'}|\\
    & \lstinline|exclusiveMinimum| & \lstinline|{type: 'boolean', default: false}| \\
    & \lstinline|exclusiveMaximum| & \lstinline|{type: 'boolean', default: false}|\\
    Must be a multiple of a given value & \lstinline|multipleOf|       & \lstinline|{type: 'number', minimum: 0, exclusiveMinimum: true}|\\
    \midrule
    \multicolumn{3}{@{}l@{}}{Schemas for type \lstinline|array|, e.g., \lstinline|\{type:'array', items: \{type: 'string'\}, minItems: 1, uniqueItems: true\}|:} \\
    \midrule
    Restrict type of all items .. & \lstinline|items|            & \lstinline|{default: {}, anyOf: [{$ref: '#'},| \\
    .. or for each individual item &                              & \lstinline| {type: 'array', minItems: 1, items: {$ref: '#'}}]}|\\
    Restrict number of items & \lstinline|minItems|         & \lstinline|{type: 'integer', minimum: 0, default: 0}|\\
    & \lstinline|maxItems|         & \lstinline|{type: 'integer', minimum: 0}|\\
    Allow items beyond those specified above & \lstinline|additionalItems|  & \lstinline|{anyOf: [{type: 'boolean'}, {$ref: '#'}], default: {}}|\\
    Enforce that items are unique & \lstinline|uniqueItems|      & \lstinline|{type: 'boolean', default: false}|\\
    \midrule
    \multicolumn{3}{@{}l@{}}{Schemas for type \lstinline|object|, e.g., \lstinline|\{type:'object', properties: \{'a':\{type: 'string'\},'b':\{enum:[0,1]\}\}\}|:} \\
    \midrule
    Specify key-value pairs allowed in the schema & \lstinline|properties|       & \lstinline|{type: 'object', additionalProperties: {$ref: '#'}, default: {}}| \\
    Restrict number of properties & \lstinline|minProperties| & \lstinline|{type: 'integer', minimum: 0, default: 0}|\\
    & \lstinline|maxProperties|    & \lstinline|{type: 'integer', minimum: 0}|\\
    Names of properties that must be present & \lstinline|required| & \lstinline|{type: 'array', items: {type: 'string'}, minItems: 1, uniqueItems: true}| \\
    Allow properties beyond those specified above & \lstinline|additional-| \lstinline|Properties| & \lstinline|{anyOf: [{type: 'boolean'}, {$ref: '#'}], default: {}}| \\
    Restrict properties whose name match a regular expression & \lstinline|pattern-| \lstinline|Properties| & \lstinline|{type: 'object', additionalProperties: {$ref: '#'}, default: {}}| \\
    When one property is present, enforce presence or type of another one & \lstinline|dependencies| & \lstinline|{type: 'object', additionalProperties: {anyOf: [ {$ref: '#'}, {type: 'array', items: {type: 'string'}, minItems: 1, uniqueItems: true}]}}|\\
    \midrule
    \multicolumn{3}{@{}l@{}}{Specifying types and combining schemas, e.g., \lstinline|\{'anyOf':[\{'type':'string'\},\{'\$ref':'\#/some/type'\}]\}|:} \\
    \midrule
    Restrict value to one type or to an array of types & \lstinline|type| & \lstinline|{anyOf: [{$ref:'#/definitions/types'}, {type: 'array', items: {$ref:'#/definitions/types'},  minItems: 1, uniqueItems: true}]}| \\
    && \lstinline|{definitions:{types:{enum:['boolean','null', 'number','integer','string','array','object']}}}| \\
    Enumerate values & \lstinline|enum| & \lstinline|{type: 'array', minItems: 1, uniqueItems: true}|\\
    Negation &  \lstinline|not|        & \lstinline|{$ref: '#'}|\\
    Conjunction &  \lstinline|allOf|      & \lstinline|{type: 'array', minItems: 1, items: {$ref: '#'}}| \\
    Disjunction & \lstinline|anyOf|      & \lstinline|{type: 'array', minItems: 1, items: {$ref: '#'}}| \\
    Exclusive or & \lstinline|oneOf|      & \lstinline|{type: 'array', minItems: 1, items: {$ref: '#'}}| \\
    
    \bottomrule
  \end{tabular}
\end{table*}

%% As JSON became more and more pervasive, developers needed a way 
%% to describe and specify how a JSON document look like. 
%% This is when JSON Schema comes to the rescue.
JSON Schema is a declarative language for defining the structure and permitted 
values of a JSON document~\cite{jsonSchema}.
JSON Schema itself uses JSON syntax.
JSON Schema is an Internet Draft from the Internet Engineering Task
Force (IETF). It is continuously evolving, currently at draft-2019-09.
This work focuses on draft04~\cite{jsonSchema4}, one of the most widely 
adopted versions of JSON Schema.

To specify which data types are allowed, JSON Schema uses the
keyword \lstinline|'type'| with one type name or a list of type
names.  For example, schema
\lstinline|{'type': 'string'}| accepts strings and schema
\lstinline|{'type': ['null', 'boolean']}| accepts null or Boolean values.
Each JSON type has a set of validation \emph{keywords} that restrict
the values a schema of this type permits.
For example, the schema \lstinline|{'type': 'integer', 'minimum': 0}|
uses the keyword \jmin{} to restrict integers to be non-negative,
while schema \lstinline|{'type': 'string', 'pattern': '^[A-Za-z0-9]+$'}|
uses the keyword \jpattern{} with a regular expression to
restrict strings to be alphanumeric.
The upper part of Table~\ref{table:jschema} lists the set of keywords 
associated with each JSON type along with their meta-schema. For
example, \jstr{} schemas have a \jminlen{} keyword with a non-negative
integer that defaults to~0.
The meta-schema acts similarly to a grammar in that it defines the
syntax for schemas, while at the same time using and thus illustrating
a realistic JSON Schema use-case.

In addition to type-specific keywords, JSON Schema allows enumerating
exact values with \enum{} and combining different schemas using a set
of logic connectives
(Table~\ref{table:jschema}, lower part).
For example, schema \lstinline|{'enum': ['a', [], 1]}| restricts the set of
permitted JSON values to the string literal \lstinline|'a'|, an empty array, or the integer~1.
Logic connectives, such as \anyof{}, \allof{}, and \jnot{}, allow schema writers to 
express disjunctions, conjunctions, and negations of schemas.
The empty schema, \lstinline|{}|, is the top of the schema hierarchy
(all documents are valid for~\lstinline|{}|). The negation of the
empty schema, \lstinline|{'not':{}}|, is the bottom of the schema
hierarchy (no documents are valid for \lstinline|{'not':{}}|.
Finally, the keyword \jref{} retrieves schemas using URIs and JSON pointers.
JSON validation against a schema with \jref{} has to satisfy the schema retrieved 
from the specified URI or JSON pointer.
We refer the interested reader to the full 
specification of JSON Schema~\cite{jsonSchema4} and its formalization~\cite{pezoa_et_al_2016}.

Given a JSON document $d$ and a JSON schema $s$, schema
validation checks if $d$ conforms to~$s$.

\begin{definition}[$\valid(d,s)$]\label{def:valid}
  For any JSON document $d$ and any JSON schema $s$,
  $\valid(d,s)\to\{\mathit{True},\mathit{False}\}$, also written
  $d:s$, decides whether $d$ is valid with respect to $s$.
\end{definition}

This decision problem $\valid(d, s)$ is shown to be PTIME-hard~\cite{pezoa_et_al_2016}
and solvable in linear time when eliminating the \juitems{} keyword as it involves sorting.
JSON validators have been implemented in most major programming languages and are widely used in several domains.

\section{Problem Statement}\label{sec:problem}

Given two schemas $s$ and $t$, our approach tries to decide whether
$s$ is a subschema of~$t$.
This section defines the subschema relation, gives concrete usage scenarios for it, and describes why deciding the subschema question is non-trivial.

\subsection{JSON Subschema}

%Assume we have two JSON schemas $s$ and $t$, and a JSON document
%$d$.
%%MP: Already explained above.
%Schema validation is a decision procedure $\valid(d,t)$ that answers the question whether
%$d : t$, i.e., whether $d$ is a valid member of $t$.
The goal of this work is to tell whether the set of documents that conform
to schema $s$ is a subset of the set of documents that
conform to schema $t$.
We aim at answering this question without enumerating all valid documents, and we call this question the \emph{JSON Subschema} decision problem.

\begin{definition}[$\sub <:$]
	\label{def:sub}
	For any two JSON schemas $s$ and $t$,
	the subschema relation, denoted $<:$, is defined as:
	$s <: t \iff \big( \forall d: \valid(d,s) \implies \valid(d,t) \big)$
\end{definition}

The relation $<:$ is a partial order, and equivalence of schemas follows directly by the anti-symmetry of the subtype relation.
\begin{definition}[$\eq \equiv$]
    \label{def:equiv}
    For any two JSON sche\-mas $s$ and $t$,
    the equivalence relation $s \equiv t$ is given by:
    $s \equiv t \iff (s <: t \wedge t <: s)$
\end{definition}
That is, two schemas are equivalent if they describe the same exact set of JSON documents.

\subsection{Usage Scenarios}

In many usage scenarios, a membership check with $\valid$ corresponds to a runtime check, whereas a subschema check provided by $\sub$ corresponds to static type checking.

\paragraph{Backward compatibility.}
One of the most pervasive use cases of JSON schemas is describing requests and responses of web
APIs. For example, version 0.6.1 of the
Washington Post \texttt{ans-schema} contains the following:

\begin{lstlisting}
'category': {
    'type': 'string',
    'enum': ['staff', 'wires', 'freelance', 'other' ]}
\end{lstlisting}

The continuous development and evolution of these APIs involves
regular changes to the corresponding JSON schemas, and developers need
to keep a close eye on such changes to avoid breaking backward
compatibility. For example, version 0.6.2 of the same schema contains:

\begin{lstlisting}
'category': {
    'type': 'string',
    'enum': ['staff', 'wires', 'freelance', 'stock', 'handout', 'other' ]}
\end{lstlisting}

Version 0.6.1 is a subschema of version 0.6.2. Assuming
the developers intend to retain backward compatibility, this evolution would
be fine for an API request argument, but it could break clients when used as
an API response.

\begin{figure}
\begin{subfigure}[t]{.38\textwidth}
\begin{lstlisting}
{'type': 'object',
 'required': ['type', 'address'],
 'properties': {
   'address': {
     'description': 'Node address',
     'type': ['string', 'null']},
   'type': {
     'description': 'Node address type; one of Hostname, ExternalIP or InternalIP',
     'type': ['string', 'null']}}}
\end{lstlisting}
\caption{}
\label{fig:kubernetes example 1}
\end{subfigure}
\hspace{1em}
\begin{subfigure}[t]{.58\textwidth}
\begin{lstlisting}
{'anyOf': [
  {'type': 'object',
    'required': ['type', 'address'],
    'properties': {
      'type': { 'enum': ['ExternalIP', 'InternalIP']},
      'address': {
        'type': 'string',
        'pattern': '^\d+\.\d+\.\d+\.\d+$'}}},
  {'type': 'object',
    'required': ['type', 'address'],
      'properties': {
        'type': { 'enum': ['Hostname']},
        'address': {
          'type': 'string',
          'pattern': '^[A-Za-z0-9.]+$'}}}]}
\end{lstlisting}
\caption{}
\label{fig:kubernetes example 2}
\end{subfigure}
\vspace*{-2mm}
\caption{Example for usage scenario of schemas as static types.}
\label{fig:static types example}
\end{figure}

\paragraph{Schemas as static types.}
JSON subschema checking can help make code more robust by flagging
some mistakes statically. Consider version 0.14.0 of the
\texttt{NodeAddress} schema from Kubernetes, extracted from its
OpenAPI specification, shown in Figure~\ref{fig:kubernetes example 1}.
With this schema, an application can check \texttt{NodeAddress} objects at runtime, but runtime errors in distributed, cloud-based systems are difficult to debug. Therefore, client code might define a stricter schema~$s$, as shown in Figure~\ref{fig:kubernetes example 2}.
Schema $s$ uses \enum{}s to constrain the values for
\mbox{\lstinline{'type'}} and \jpattern{}s to constrain the values for
\lstinline{'address'}, with an \anyof{} to provide two cases.  By
being stricter, schema $s$ can rule out more bugs. The static
check \mbox{$s<:\texttt{NodeAddress}$} can validate that $s$
is indeed a subschema.

%% \begin{figure}
%%   \centerline{\includegraphics[width=\columnwidth]{figures/pipeline_sklearn.png}}
%%   \vspace*{-2mm}
%%   \caption{\label{fig:pipeline_sklearn}AI pipeline with slow dataset format exception.}
%% \end{figure}

%% \begin{figure}
%%   \centerline{\includegraphics[width=\columnwidth]{figures/pipeline_lale_1.png}}
%%   \centerline{\includegraphics[width=\columnwidth]{figures/pipeline_lale_2.png}}
%%   \vspace*{-2mm}
%%   \caption{\label{fig:pipeline_lale}AI pipeline with fast dataset schema exception.}
%% \end{figure}

\paragraph{Machine learning pipelines.}

Machine-learning pipelines are of little use if the data is formatted
incorrectly~\cite{Breck2019}. For example, here is the input schema
for the NMF transformer from scikit-learn~\cite{buitinck_et_al_2013},
which performs non-negative matrix factorization.

\begin{lstlisting}
{   'type': 'array',
    'items': {
        'type': 'array',
        'items': {
            'type': 'number', 'minimum': 0.0}}}}}
\end{lstlisting}

The outer array ranges over rows (samples in machine learning) and the
inner array ranges over columns (features in machine learning). NMF
requires non-negative numbers as input, captured by the keyword \jmin.
This transformer is typically used in the middle of a larger
machine-learning pipeline, after data cleansing transformers but
before a classifier or regressor.
This schema can check a specific part of a dataset at a specific point
in the pipeline, but this check has to be repeated for production
data, which may differ from training data for various reasons.

Instead, several sources argue that the dataset itself should also be
associated with a schema~\cite{Breck2019,campbell_2017,sculley_et_al_2015}. For example,
here is a schema for the features of the California Housing dataset.

\begin{lstlisting}
{   'type': 'array',
    'items': {
        'type': 'array', 'minItems': 4, 'maxItems': 4,
        'items': [
        {   'description': 'AveRooms',   'type': 'number', 'minimum': 0.0},
        {   'description': 'Population', 'type': 'number', 'minimum': 0.0},
        {   'description': 'Latitude',   'type': 'number', 'minimum': 0.0},
        {   'description': 'Longitude',  'type': 'number'}]}}
\end{lstlisting}

The outer array ranges over rows, each one describing a different
district in California. The inner array ranges over columns, each one
describing a feature that might be helpful for predicting the median
house value in that district. Instead of a single schema for all
items, there are separate per-item schemas (see the meta-schema for
the \jitems{} keyword in Table~\ref{table:jschema}).

Is the dataset compatible with the input expected by the NMF transformer?
California is north of the equator, so the latitude is positive, and west of
Greenwich, so the longitude is negative. Due to the negative
longitudes (expressed by the absence of the \texttt{minimum} keyword), this schema is not a subschema of the input schema of NMF.
Machine learning engineers could add 360 degrees to the longitude and
change its schema to
\lstinline|{'description': 'Longitude', 'type': 'number', 'minimum': 0.0}|.
Then, the dataset schema would be a subschema of the NMF input schema,
thanks to the keywords
\lstinline|'minItems': 4, 'maxItems': 4|
which guarantee the absence of additional columns with unspecified
schemas.  Later, when this model is used in production, the schema can
be used to check the validity of new samples. For instance, if there
is a sample that comes not from California but from south of the
equator, a schema check could catch the negative latitude early and
explain which assumption it violated.

%% \subsection{Examples}

%% \begin{alltt}\textcolor{red}{TODO}\scriptsize
%% - expanded version of Figure~\ref{fig:galois} that includes
%%   a minimal example, as suggested by Michael:
%%   show two simple schemas that are subtypes and that need a bit of
%%   canonicalization and some very simple extraction of fragments
%% \end{alltt}

\subsection{Challenges}\label{subsec:challenges}

\begin{figure*}
\begin{subfigure}[b]{.3\textwidth}
\begin{lstlisting}
{'type': [
     'null',
     'string'], 
 'not': {
     'enum': ['']}}
\end{lstlisting}
\vspace*{-2mm}
\caption{}
\label{fig:str1}
\end{subfigure}
\hfill
\begin{subfigure}[b]{.3\textwidth}
\begin{lstlisting}
{'anyOf': [
     {'type': 'null'},
     {'type': 'string'}],
 'not': {
     'type': 'string',
     'enum': ['']}}
\end{lstlisting}
\vspace*{-2mm}
\caption{}
\label{fig:str2}
\end{subfigure}
\hfill
\begin{subfigure}[b]{.33\textwidth}
%\begin{lstlisting}
%{'anyOf': [{'type': 'string'}, {'type': 'null'}],
% 'anyOf': [{'type': 'array'}, {'type': 'object'},
%           {'type': 'number'}, {'type': 'integer'},
%           {'type': 'boolean'}{'type': 'null'},
%           {'type': 'string', 'pattern': '.+'}]}
%\end{lstlisting}
\begin{lstlisting}
{'allOf': [
    {'anyOf': [
        {'type': 'null'}, 
        {'type': 'string'}]},
    {'not': {
        'type': 'string',
        'enum': ['']}}]}
\end{lstlisting}
\vspace*{-2mm}
\caption{}
\label{fig:str3}
\end{subfigure}
\vfill
\begin{subfigure}[b]{.53\textwidth}
%\begin{lstlisting}
%{'allOf': 
%  [{'anyOf': [{'type': 'string'}, {'type': 'null'}]},
%   {'anyOf': [{'type': 'array'}, {'type': 'object'}, {'type': 'number'},
%              {'type': 'integer'}, {'type': 'boolean'}{'type': 'null'},
%              {'type': 'string', 'pattern': '.+'}]}]}
%\end{lstlisting}
\begin{lstlisting}
{'allOf': [
    {'anyOf': [
        {'type': 'null'},
        {'type': 'string'}]},
    {'anyOf': [
        {'type': 'boolean'}, {'type': 'null'}, 
        {'type': 'number'}, {'type': 'integer'},
        {'type': 'array'}, {'type': 'object'},
        {'type': 'string', 'pattern': '.+'}]}]}
\end{lstlisting}
\vspace*{-2mm}
\caption{}
\label{fig:str4}
\end{subfigure}
\hfill
\begin{subfigure}[b]{.45\textwidth}
\begin{lstlisting}
{'anyOf': [
    {'type': 'null'},
    {'type': 'string', 'pattern': '.+'}]}
\end{lstlisting}
%    'minLength': 0, 'maxLength': (*$\infty$*)},
\vspace*{-2mm}
\caption{}
\label{fig:str5}
\end{subfigure}
\vspace*{-2mm}
\caption{Five syntactically different but semantically equivalent schemas for a value that is either a non-empty string or null.}
\label{fig:str}
\end{figure*}

JSON schemas define the nested \emph{structure} and \emph{valid values}
permitted in a set of JSON documents.  The rich feature set of JSON
Schema makes establishing or refuting a subschema relationship
between two schemas non-trivial.
Even for simple, structurally similar schemas, such as
\lstinline|{'enum': [1, 2]}| and 
\lstinline|{'enum': [2, 1]}|,
equivalence does not hold through textual equality.
There are several challenges for algorithmically checking the subtype relation of JSON schemas.

First, the schema language is flexible and the same set of JSON values, i.e., the same type, 
could be described in several different syntactical forms, i.e., schemas. 
For example, Figure~\ref{fig:str} shows five equivalent schemas describing a JSON value 
that is either a non-empty string or null.

Second, even for primitive types, such as strings and numbers, nominal subtyping 
is not applicable. 
JSON Schema lets users specify various constraints on primitive types, resulting in non-trivial interactions that are not captured by nominal types.
For example, one cannot infer that an \jint{} schema is a subtype of a \jnum{}
schema without properly comparing the range and multiplicity constraints of the schemas.

Third, logic connectives combine non-homogeneous types
such as \lstinline{'string'} and \lstinline{'null'} in
Figure~\ref{fig:str2}.
Moreover, enumerations restrict types to predefined values, which require careful handling, especially when enumerations interact with non-enumerative types, such as in Figures~\ref{fig:str1}, \ref{fig:str2}, and~\ref{fig:str3}.

Fourth, the schema language allows implicit conjunctions and disjunctions.
For example, Figure~\ref{fig:str2} has an implicit top-level conjunction
between the subschemas under \anyof{} and \jnot.
As another example, a schema that lacks a type keyword, such as
\lstinline|{'pattern': '.+'}|, has an implicit disjunction between all
possible types, while still enforcing any type-specific keyword, such
as the pattern for strings only. For illustration purposes,
Figure~\ref{fig:str4} makes this implicit disjunction explicit.

Finally, JSON Schema also allows uninhabited types.
That is, a schema can be syntactically valid yet semantically
self-contra\-dicting, e.g.,
\lstinline|{'type':'number','minimum':5,'maximum':0}|.
Such schemas validate no JSON value at all and complicate reasoning about subtyping.

\section{Algorithm}
\label{sec:algo}

This section describes how we address the problem of checking whether one JSON schema is a subtype of another.
Because JSON schemas are complex, creating a subtype checker for arbitrary schemas directly would necessitate a complex algorithm to handle all of its variability.
Instead, we decompose the problem into three steps.
The first step canonicalizes a given schema into an equivalent but more standardized schema  (Section~\ref{sec:canonicalization}).
The second step further simplifies a schema by eliminating
enumerations, negation, intersection, and union of schemas where
possible (Section~\ref{sec:simplification}).
Table~\ref{table:cansimp} summarizes the first two steps.
Finally, the third step checks for two canonicalized and simplified schemas whether one is a subtype of the other by extracting and comparing type-homogeneous schema fragments (Section~\ref{sec:subtyping}).

\begin{table*}\small
  \caption{\label{table:cansimp}Canonicalization and simplification guarantees.}
  \texttt{\begin{tabular}{@{}p{4em}p{32mm}p{42mm}p{42mm}@{}}
    \toprule
    \textrm{\textbf{Language}} & \multicolumn{3}{c}{\textrm{\textbf{Use of feature in schemas}}} \\
    \cmidrule{2-4}
    \raisebox{1mm}{\textrm{\textbf{feature}}} & \textrm{\textbf{Full JSON schema}} & \textrm{\textbf{Canonicalized}} & \textrm{\textbf{Simplified}}\\
    \midrule
    null    & \textrm{Yes} & \textrm{Yes} & \textrm{Yes}\\
    boolean & \textrm{Yes}
    & \textrm{Represented as} enum & \textrm{Represented as} enum\\
    string  & \textrm{All keywords}
    & \textrm{Keyword} pattern \textrm{only} & \textrm{Keyword} pattern \textrm{only}\\
    number  & \textrm{All keywords}
    & \textrm{All keywords}
    & \textrm{All keywords} \\
    integer & \textrm{All keywords}
    & \textrm{Eliminated} & \textrm{Eliminated}\\
    array   & \textrm{All keywords}
    & \textrm{All keywords; \mbox{items} is always list}
    & \textrm{All keywords; \mbox{items} is always list}\\
    object  & \textrm{All keywords}
    & \mbox{\textrm{Only} \{min,max\}Properties,} patternProperties, required \textrm{keywords}
    & \mbox{\textrm{Only} \{min,max\}Properties,} patternProperties, required \textrm{keywords}\\
    \midrule
    enum    & \textrm{Heterogeneous, any type}
    & \textrm{Homogeneous, any type} & \textrm{Only for \texttt{boolean}}\\
    not     & \textrm{Multiple connectives}
    & \textrm{Single connective} & \textrm{Only for \texttt{number}, \texttt{array}, \texttt{object}}\\
    allOf   & \textrm{Multiple connectives}
    & \textrm{Single connective} & \textrm{Only for \texttt{not}}\\
    anyOf   & \textrm{Multiple connectives}
    & \textrm{Single connective} & \textrm{Only for \texttt{not}, \texttt{allOf}, \texttt{array}, \texttt{object}, and disjoint \texttt{number}.}\\
    oneOf   & \textrm{Multiple connectives}
    & \textrm{Eliminated} & \textrm{Eliminated}\\
    \bottomrule
  \end{tabular}}
\end{table*}

\paragraph{Notation}
We formalize canonicalization and simplification via rewrite
rules of the form $s_1->s_2$.
%% MP: These aren't rules, and we already introduce <: in Def. 3.1
%, and formalizes subschema checking via rules of the form $s_1<:s_2$.
The notation $s.k$ indicates access of property $k$ in schema $s$.
For any JSON schema $s$, helper function $\dom(s)$ returns its
property names, i.e., the set of keys in the key-value map~$s$.
The notation $s[k |->v ]$ indicates a substitution, which is a copy of
$s$ except that the mapping of key~$k$ is added or changed to value~$v$.
The notation $\delim{[}\ldots\delim{]}$ indicates a JSON array and
the notation $\delim{\{}\ldots\delim{\}}$ indicates a JSON object.
The notation $\delim{\{}k\punct{:}v\mid\ldots\delim{\}}$ indicates a JSON
object comprehension.
The notation $a \parallel b$ is a default operator that returns
$a$ if it is defined and $b$ otherwise.

\subsection{JSON Schema Canonicalization}\label{sec:canonicalization}

%Therefore, to tackle the subschema problem, we devise a syntactic approach that borrows 
%concepts of \emph{structural subtyping}.
%In structural type systems, types are related based on the structure
%and types of their members~\cite{Cardelli1988}.

This section introduces a canonicalization procedure
that compiles any JSON schema into an equivalent canonical schema.
Column ``Canonicalized'' of Table~\ref{table:cansimp} summarizes the
properties that the canonicalizer establishes.
Given any JSON schema as input, canonicalization terminates
and produces a semantically equivalent, canonical JSON schema as
output.

The canonicalization enforces two main simplifications.
First, JSON Schema allows schemas to mix specifications of
different types.  To enable local, domain-specific reasoning in the
subtype checker, canonicalization splits up these schemas into
smaller, homogeneously typed schemas that are combined using
logic connectives.
Second, JSON Schema allows many alternative ways to represent the
same thing. Additionally, there are keywords that can be omitted
and the defaults are then assumed.  Canonicalization picks, when
possible, one form, and supplies omitted defaults explicitly.

\begin{figure*}\small
    ~
    \inference[multiple types~]
    {s.\type = \delim{[}\tau_1, \dots, \tau_n\delim{]}}
    {s -> \delim{\{}\anyof \punct{:} \delim{[}s[\type |-> \tau_1], \dots, s[\type |-> \tau_n]\delim{]\}}}
    \vspace{1em}
    \inference[multiple connectives~]
    {\dom(s)\cap\{\enum,\anyof,\allof,\oneof,\jnot\}\neq\emptyset
     & \dom(s)\setminus\{c\}\neq\emptyset}
    {s -> \delim{\{}\allof \punct{:} \delim{[\{}c \punct{:} s.c\delim{\}}, \delim{\{}k \punct{:} s.k \;\arrowvert\;  k\in(\dom(s)\setminus\{c\})\delim{\}]\}}}
    \vspace{1em}
    \inference[missing type~]
    {\dom(s) \cap \{\type, \enum, \anyof, \allof, \oneof, \jnot\} = \emptyset}
    {s -> s[\type |-> \jtypes]}
    \caption{Non-type-specific canonicalization rules that ensure exactly one type, enum, or logic connective.}
    \label{rules:multicon}
\end{figure*}

Figure~\ref{rules:multicon} presents non-type-specific
canonicalization rules whose purpose it is to enable reasoning about
one type or connective at a time.
Rule \emph{multiple types} applies to schemas whose type is a list,
such as in the example in Figure~\ref{fig:str1}, making the implicit
disjunction explicit using \anyof, as shown in Figure~\ref{fig:str2}.
Rule \emph{multiple connectives} applies to schemas that contain a
connective mixed with other connectives, such as in the example in
Figure~\ref{fig:str2}, making the implicit conjunction explicit using
\allof, as shown in Figure~\ref{fig:str3}.
Rule \emph{missing type} generously assumes all JSON types are
possible, yielding an implicit disjunction to be further canonicalized
by the multiple-types rule.

\begin{figure*}\small
    ~
    \inference[missing keyword~]
    {s.\type=\tau & \tau\in\jtypes & \tau\neq\jstr & k\in\kw(\tau) & k\notin\dom(s)}
    {s -> s[k |-> \default(k)]}
    \vspace{1em}
    \inference[irrelevant keywords~]
    {s.\type = \tau & \tau \in \jtypes}
    {s -> \delim{\{}k \punct{:} s.k \;\arrowvert\; k \in (\mathit{dom}(s) \cap (\kw(\tau) \cup \{\type,\enum\})\delim{\}}}
    \vspace*{1em}
    \inference[integer~]
    {s.\type = \jint}
    {s -> s[\type |-> \jnum, \jmulof |-> \textit{lcm}(1, s.\jmulof \parallel 1)]}
    \vspace{1em}
    \inference[heterogeneous enum~]
    {s.\enum = \delim{[}v_1, \dots, v_n\delim{]} & \exists j, \typeof(v_j)\neq\typeof(v_1)}
    {\begin{array}{@{}l@{}l@{}}
      s -> \delim{\{}\anyof\punct{:}\delim{[}
      & s[\enum |-> \delim{[}v_i\mid\typeof(v_i)=\typeof(v_1)\delim{]}, \type |-> \typeof(v_1)],\\
      & s[\enum |-> \delim{[}v_j\mid\typeof(v_j)\neq\typeof(v_1)\delim{]}]\delim{]\}}
    \end{array}}
    \vspace{1em}
    \inference[oneOf~]
    {s.\oneof = \delim{[}s_1, \dots, s_n\delim{]}}
    {\begin{array}{l@{}l}
      s -> \delim{\{}\anyof\punct{:}\delim{[}
      & \delim{\{}\allof\punct{:}\delim{[}s_1, \delim{\{}\jnot \punct{:} s_2\delim{\}}, \dots, \delim{\{}\jnot \punct{:} s_n\delim{\}]\}},\\
      & \dots,\\
      & \delim{\{}\allof \punct{:} \delim{[\{}\jnot \punct{:} s_1\delim{\}}, \dots, \delim{\{}\jnot \punct{:} s_{n-1}\delim{\}}, s_n\delim{]\}]\}}
    \end{array}}
    \vspace{1em}
  \inference[string without maxlength~]
  {s.\type=\jstr & s.\jpattern=p & s.\jminlen=a & \jmaxlen\notin\dom(s)}
  {s -> \delim{\{}\type\punct{:}\jstr, \jpattern\punct{:}p\cap\texttt{\textquotesingle\string^.\{}a\texttt{\}\textquotesingle}\delim{\}}}
  \vspace*{1em}
  \inference[string with maxlength~]
  {s.\type=\jstr & s.\jpattern=p & s.\jminlen=a & s.\jmaxlen=b}
  {s -> \delim{\{}\type\punct{:}\jstr, \jpattern\punct{:}p\cap\texttt{\textquotesingle\string^.\{}a,b\texttt{\}\$\textquotesingle}\delim{\}}}
    \vspace*{1em}
  \inference[array with one schema for all items~]
  {s.\type=\jarray & s.\jitems=\delim{\{}\ldots\delim{\}}}
  {s -> s[\jitems |-> \delim{[]}, \jadditems |-> s.\jitems]}
  \vspace*{1em}
  \inference[array with additionalItems false~]
  {s.\type=\jarray & s.\jadditems=\texttt{false}}
  {s -> s[\jadditems |-> \delim{\{}\jnot\punct{:}\delim{\{\}\}}]}
  \vspace*{1em}
  \inference[object with additionalProperties false~]
  {s.\type=\jobj & s.\jaddp=\texttt{false}}
  {s -> s[\jaddp \mapsto \delim{\{}\jnot\punct{:}\delim{\{\}\}}]}
  \vspace{1em}
  \inference[\parbox{12mm}{object with properties}~]
  { s.\type=\jobj
    & s.\jprops=\delim{\{}k_1\punct{:}s_{k_1},\ldots,k_n\punct{:}s_{k_n}\delim{\}}\\
    s.\jaddp=\delim{\{}\ldots\delim{\}}
    & s.\jpatp=\delim{\{}p_1\punct{:}s_{p_1},\ldots,p_m\punct{:}s_{p_m}\delim{\}}}
  { \begin{array}{@{}l@{}l@{}l@{}}
      s -> &
        s[\jpatp |-> \delim{\{}
          & \texttt{\textquotesingle\string^}k_1\texttt{\$\textquotesingle} \punct{:}s_{k_1},
            \quad\ldots,\quad
            \texttt{\textquotesingle\string^}k_n\texttt{\$\textquotesingle} \punct{:}s_{k_n},\\
        & & p_1\setminus\texttt{\textquotesingle\string^(}k_1\texttt{|}\textrm{...}\texttt{|}k_n\texttt{)\$\textquotesingle} \punct{:} s_{p_1},
            \quad\ldots,\quad
            p_m\setminus\texttt{\textquotesingle\string^(}k_1\texttt{|}\textrm{...}\texttt{|}k_n\texttt{)\$\textquotesingle} \punct{:} s_{p_m},\\
        & & \neg\texttt{\textquotesingle\string^(}k_1\texttt{|}\textrm{...}\texttt{|}k_n\texttt{)\$|}p_1\texttt{|}\textrm{...}\texttt{|}p_m\texttt{\textquotesingle} \punct{:} s.\jaddp\delim{\}}]\\
        & \multicolumn{2}{@{}l}{\setminus\{\jprops,\jaddp\}}
    \end{array}}
  \vspace*{1em}
  \inference[\parbox{20mm}{object with string list dependencies}~]
  {s.\type=\jobj & s.\jdeps=\delim{\{}k_i\punct{:}\delim{[}k_{i_1},\ldots,k_{i_n}\delim{]\}}\cup d_\textit{rest}}
  {s -> s[\jdeps |-> d_\textit{rest} \cup \delim{\{}k_i : \delim{\{}
          \type \punct{:} \jobj,
          \jreq \punct{:} \delim{[}k_{i_1},\ldots,k_{i_n}\delim{]\}\}}]}
  \vspace*{1em}
  \inference[\parbox{20mm}{object with sche\-ma dependencies}~]
  {s.\type=\jobj & s.\jdeps=\delim{\{}k_i\punct{:}s_i\delim{\}}\cup d_\textit{rest}}
  {\begin{array}{@{}l@{}l@{}}
     s ->\delim{\{}\allof \punct{:}\delim{[}
     & s[\jdeps |-> d_\textit{rest}],\\
     & \delim{\{}\anyof\punct{:}\delim{[} s_i,
         \delim{\{}
           \type \punct{:}\jobj,
           \jprops \punct{:} \delim{\{}k_i\punct{:}\jnot\punct{:}\delim{\{\}\}\}} \delim{]\}\}}\\
   \end{array}}
   \vspace*{1em}
  \inference[\parbox{26mm}{object with overlapping patternProperties}~]
  {s.\type=\jobj & s.\jpatp=\delim{\{}p_i\punct{:}s_i\delim{\}}\cup
    \delim{\{}p_j\punct{:}s_j\delim{\}} \cup d_\textit{rest}}
  {\begin{array}{@{}l@{}l@{}}
     s -> s[&\jpatp |-> \\
     &\delim{\{}p_i\cap p_j\punct{:}\delim{\{}\allof \punct{:}\delim{[}s_i,s_j\delim{]\}\}}
     \cup \delim{\{}p_i\cap \neg p_j\punct{:}s_i\delim{\}}
     \cup \delim{\{}\neg p_i\cap p_j\punct{:}s_j\delim{\}}
     \cup d_\textit{rest}
     ]     
   \end{array}}
  \caption{\label{rules:canonical}Type-specific canonicalization rules.}
\end{figure*}

Figure~\ref{rules:canonical} presents type-specific canonicalization
rules whose purpose it is to reduce the number of cases to handle for
later simplification and subschema rules.
Rule \emph{missing keyword} adds the default for a keyword if there is
a single type and the keyword for that type is missing, using a helper
function $\default$ that returns the default from the meta-schema in
Table~\ref{table:jschema} and maps $\jmin$ to $-\infty$ and $\jmax$ to
$\infty$ for convenience.
Rule \emph{irrelevant keywords} strips out spurious keywords that do
not apply to a given type (or to any type), using a helper function
$\kw$ that returns the relevant keywords in Table~\ref{table:jschema}.
Rule \emph{integer} rewrites integer schemas to number schemas with
the appropriate $\jmulof$.
Rule \emph{heterogeneous enum}, when iterated, ensures that each
enumeration contains only values from a single type, using a helper
function $\typeof$ that maps a concrete JSON value to its type.
Rule \emph{oneOf} eliminates the $\oneof$ keyword by rewriting the
exclusive or into a disjunction of conjunctions.

The two \emph{string} rules eliminate the keywords $\jminlen$ and
$\jmaxlen$ by compiling them into regular expressions, so that after
canonicalization, string schemas only have the keyword $\jpattern$.

The two \emph{array} rules handle keywords that can be specified in
multiple different ways, as indicated by meta-schemas with $\anyof$ in
Table~\ref{table:jschema}.
Rule \emph{array with one schema for all items} changes the keyword
$\jitems$ from a single schema to a list (empty) of schemas, by moving
the schema into $\jadditems$.
Rule \emph{array with additionalItems false} changes the keyword
$\jadditems$ from a boolean to a schema (bottom).

The five \emph{object} rules eliminate the object keywords $\jprops$,
$\jaddp$, and $\jdeps$ by rewriting them into the keywords $\jreq$ and
$\jpatp$, and ensure that $\jpatp$ uses non-overlapping regular
expressions. The object rules are the most intricate out of the
canonicalization rules because in JSON Schema, object schemas have the
largest number of special cases. Reducing the cases reduces the
complexity of subsequent rules for simplification and subschema
checking.
Rule \emph{object with additionalProperties false} changes the keyword
$\jaddp$ from a boolean to a schema (bottom).
Rule \emph{object with properties} eliminates $\jprops$ and $\jaddp$
by rewriting them into $\jpatp$. First, it turns each property key
$k_i$ from $\jprops$ into a pattern property with the regular
expression \mbox{$\texttt{\textquotesingle\string^}k_i\texttt{\$\textquotesingle}$} that accepts
exactly~$k_i$. Second, it subtracts all keys $k_i$ from each of the
original pattern properties so they only apply as a fall-back, where
the notation $p_1\setminus p_2$ indicates regular expression
subtraction. Third, it creates a tertiary fall-back regular expression
that applies when neither the original properties nor the original
pattern properties match, using the notation $\neg p$ for the
complement of a regular expression, and associates that regular
expression with the original $\jaddp$.  Finally, it removes the
eliminated keywords $\jprops$ and $\jaddp$ from the resulting schema.
Rule \emph{object with string list dependencies}, when iterated,
eliminates dependencies specified as a list of property names by
rewriting them into dependencies specified as a schema.
Rule \emph{object with schema dependencies}, when iterated, eliminates
dependencies of a key $k_i$ on a schema $s_i$ by rewriting them into a
conjunction with either $s_i$ or a schema that enforces the absence
of~$k_i$.
Rule \emph{object with overlapping pattern properties} rewrites a pair
of pattern properties with overlapping regular expressions $p_i$ and
$p_j$ so their regular expressions match only disjoint keys, by
replacing them with different schemas for the cases where (i)~both
$p_i$ and $p_j$ match, (ii)~only $p_i$ and not $p_j$ matches, and
(iii)~only $p_j$ and not $p_i$ matches. When iterated, this eliminates
all overlapping patterns, facilitating local reasoning.

%\mpx{An example for the object canonicalization rules would be useful (if we have space).}

% \paragraph{Termination}

\subsection{Simplification of Enumeration, Negation, Intersection, and Union Types}
\label{sec:simplification}

\begin{figure*}\small
    ~
    \inference[multi-valued enum~]
    {s.\type = \tau & \tau \neq \jbool & s.\enum = \delim{[}v_1,\ldots,v_n\delim{]} & n>1}
    {s -> \delim{\{}\anyof\punct{:}\delim{[\{}\type\punct{:}\tau, \enum\punct{:}\delim{[}v_1\delim{]\}},\ldots,\delim{\{}\type\punct{:}\tau, \enum\punct{:}\delim{[}v_n\delim{]\}]\}}}
    \vspace{1em}
    \inference[null enum~]
    {s.\type=\jnull & s.\enum=\delim{[}\jnull\delim{]}}
    {s -> \delim{\{}\type\punct{:}\jnull\delim{\}}}
    \hspace{1em}
    \inference[string enum~]
    {s.\type=\jstr & s.\enum=\delim{[}v\delim{]}}
    {s -> \delim{\{}\type\punct{:}\jstr, \jpattern\punct{:}\texttt{\textquotesingle\string^}v\texttt{\$\textquotesingle}\delim{\}}}
    \vspace{1em}
    \inference[number enum~]
    {s.\type=\jnum & s.\enum=\delim{[}v\delim{]}}
    {s -> \delim{\{}\type\punct{:}\jnum, \jmin\punct{:}v, \jmax\punct{:}v\delim{\}}}
    \vspace{1em}
    \inference[array enum~]
    {s.\type=\jarray & s.\enum=\delim{[[}v_1,\ldots,v_n\delim{]]}}
    {s -> \delim{\{} \type\punct{:}\jarray, \jminitems\punct{:}n, \jmaxitems\punct{:}n,\jitems\punct{:}\delim{[\{}\enum\punct{:}\delim{[}v_1\delim{]\}},\ldots,\delim{\{}\enum\punct{:}\delim{[}v_n\delim{]\}]}}
    \vspace{1em}
    \inference[object enum~]
    {s.\type=\jobj & s.\enum=\delim{[\{}k_1\punct{:}v_1,\ldots,k_n\punct{:}v_n\delim{\}]}}
    {\begin{array}{l@{}l}
      s -> \delim{\{}
      & \type\punct{:}\jobj, \jreq\punct{:}\delim{[}k_1,\ldots,k_n\delim{]},\jaddp\punct{:}\jfalse,\\
      & \jprops\punct{:}\delim{\{}k_1\punct{:}\delim{\{}\enum\punct{:}\delim{[}v_1\delim{]\}},\ldots,k_n\punct{:}\delim{\{}\enum\punct{:}\delim{[}v_n\delim{]\}\}\}}
    \end{array}}
    \caption{Simplification rules to eliminate $\enum$ except for type \texttt{boolean}.}
    \label{rules:elim_enum}
\end{figure*}

This section describes the second step of our approach: a simplifier that compiles any canonical JSON schema into an equivalent simplified schema.  Column ``Simplified'' of
Table~\ref{table:cansimp} summarizes the properties that the
simplifier establishes. The simplifier eliminates many cases of
$\enum$, $\jnot$, $\allof$, and $\anyof$ connectives, thus making
subschema checking rules less complicated.  Unfortunately, in some
cases, JSON schema cannot express schemas without these connectives,
so they cannot be completely simplified away.

Figure~\ref{rules:elim_enum} shows simplification rules for $\enum$,
which turn schemas with enums into schemas without enums by using
restrictions keywords from their corresponding types instead.
Rule \emph{multi-valued enum} puts each non-Boolean enumerated value
into an $\enum$ of its own.
The rules for primitive types ($\jnull$, $\jstr$, and $\jnum$) express
a primitive enumerated value via a schema that does not use an
$\enum$. For instance, in Figure~\ref{fig:str3}, the enumerated empty
string value is compiled into the regular expression \lstinline|'^$'|
before computing its complement \lstinline|'.+'| in
Figure~\ref{fig:str4}.
The rules for structured types ($\jarray$ and $\jobj$) push down
$\enum$s to components; with iteration, the rules eventually reach
primitive types and the $\enum$s get eliminated.
The simplifier does not eliminate Boolean enumerations as the space of
values is finite and there is no other way to specify the true or
false value.

\begin{figure*}\small
    ~
    \inference[not type~]
    {s.\type = \tau & \tau \in \jtypes}
    {\delim{\{}\jnot\punct{:}s\delim{\}} -> \delim{\{} \anyof \punct{:} \delim{[} \neg s, \delim{\{} \type \punct{:} (\jtypes \setminus \tau) \delim{\}]\}}}
    \hspace{1em}
    \inference[complement null~]
    {s.\type=\jnull}
    {\neg s -> \delim{\{}\jnot\punct{:}\delim{\{\}\}}}
    \vspace{1em}
    \inference[\parbox{14mm}{complement boolean}~]
    {s.\type=\jbool & s.\enum=e}
    {\neg s -> \delim{\{}\type\punct{:}\jbool, \enum\punct{:}\neg e\delim{\}}}
    \hspace{1em}
    \inference[\parbox{14mm}{complement string}~]
    {s.\type=\jstr & s.\jpattern=p}
    {\neg s -> \delim{\{}\type\punct{:}\jstr, \jpattern\punct{:}\neg p\delim{\}}}
    %% %
    %% \vspace{1em}
    %% \inference[\parbox{13mm}{complement number}~]
    %% {s.\type=\jnum & \jmulof\notin\dom(s)}
    %% {\begin{array}{@{}l@{}l@{}}
    %%   \neg s -> \delim{\{} & \anyof\punct{:}\delim{[}\\
    %%   & \quad\delim{\{}\type\punct{:}\jnum, \jmin\punct{:}s.\jmax, \jxmin\punct{:}s.\jxmax\delim{\}},\\
    %%   & \quad\delim{\{}\type\punct{:}\jnum, \jmax\punct{:}s.\jmin, \jxmax\punct{:}s.\jxmin\delim{\}]\}}
    %% \end{array}}
    %
    \vspace{1em}
    \inference[not anyOf~]
    {s=\delim{\{}\anyof\punct{:}\delim{[}s_1,\ldots,s_n\delim{]\}}}
    {\delim{\{}\jnot\punct{:}s\delim{\}} ->
%    {\neg s ->
       \delim{\{}\allof\punct{:}\delim{[\{}\jnot\punct{:}s_1\delim{\}},\ldots,\delim{\{}\jnot\punct{:}s_n\delim{\}]\}}}
    \vspace{1em}
    \inference[not allOf~]
    {s = \delim{\{}\allof\punct{:}\delim{[}s_1,\ldots,s_n\delim{]\}}}
    {\delim{\{}\jnot\punct{:}s\delim{\}} ->
%    {\neg s ->
       \delim{\{}\anyof\punct{:}\delim{[\{}\jnot\punct{:}s_1\delim{\}},\ldots,\delim{\{}\jnot\punct{:}s_n\delim{\}]\}}}
    \hspace{1em}
    \inference[not not~]
    {s = \delim{\{}\jnot\punct{:}s_1\delim{\}}}
    {\delim{\{}\jnot\punct{:}s\delim{\}} ->
%    {\neg s ->
      s_1}    
    \caption{Simplification rules to eliminate negation, except for types \texttt{number}, \texttt{array}, and \texttt{object}.}
    \label{rules:elim_not}
\end{figure*}

Figure~\ref{rules:elim_not} shows simplification rules for $\jnot$, which eliminates negation except for numbers, arrays, and objects.
Rule \emph{not type} turns a schema of a given type $\tau$ into a
disjunction of either $\neg s$ (the complement of the values permited
by $s$ in $\tau$) or values of any type other than~$\tau$. An example
for this rule in action is the rewrite from Figure~\ref{fig:str3} to
Figure~\ref{fig:str4}, where the complement of a string schema
introduces schemas of all non-string types.
The \emph{complement} rules for $\jnull$, $\jbool$, and $\jstr$ use
the bottom type $\delim{\{}\jnot\punct{:}\delim{\{\}\}}$, the complement of the Boolean enumeration, and the complement
of the regular expression, respectively.
Rules \emph{not anyOf} and \emph{not allOf} use De Morgan's
theorem to push negation through disjunction and conjunction, and rule
\emph{not not} eliminates double negation.
Unfortunately, JSON Schema is not closed under complement for numbers, arrays, and objects. For example, the complement of schema
\lstinline|{type:'number',multipleOf:1}| is
$\mathbb{R}\setminus\mathbb{Z}$, which cannot be expressed in JSON
schema without a negation. Similar counter-examples exist for array
and object schemas. The case of negated number schemas is handled
later during subschema checking.

\begin{figure*}\small
    ~
    \inference[singleton allOf~]
    {s.\allof = \delim{[}s_1\delim{]}}
    {s -> s_1}
    \hspace{1em}
    \inference[fold allOf~]
    {s.\allof = \delim{[}s_1,s_2,\ldots,s_n\delim{]} & n \geq 3}
    {s -> \delim{\{}\allof\punct{:}\delim{[}s_1, \delim{\{}\allof\punct{:}\delim{[}s_2,\ldots,s_n\delim{]\}]\}}}
    \vspace{1em}
    \inference[\parbox{18mm}{intersect hetero\-geneous types}~]
    {s_1.\type \neq s_2.\type}
    {\delim{\{}\allof\punct{:}\delim{[}s_1,s_2\delim{]\}} -> \delim{\{}\jnot\punct{:}\delim{\{\}\}}}
    \hspace{1em}
    \inference[intersect null~]
    {s_1.\type=\jnull & s_2.\type=\jnull}
    {\delim{\{}\allof\punct{:}\delim{[}s_1,s_2\delim{]\}} -> \delim{\{}\type\punct{:}\jnull\delim{\}}}
    \vspace{1em}
    \inference[intersect boolean~]
    {s_1.\type=\jbool & s_2.\type=\jbool}
    {\delim{\{}\allof\punct{:}\delim{[}s_1,s_2\delim{]\}} -> \delim{\{}\type\punct{:}\jbool, \enum\punct{:}s_1.\enum\cap s_2.\enum\delim{\}}}
    \vspace{1em}
    \inference[intersect string~]
    {s_1.\type=\jstr & s_2.\type=\jstr}
    {\delim{\{}\allof\punct{:}\delim{[}s_1,s_2\delim{]\}} -> \delim{\{}\type\punct{:}\jstr, \jpattern\punct{:}s_1.\jpattern\cap s_2.\jpattern\delim{\}}}
    \vspace{1em}
    \inference[\parbox{8mm}{intersect \mbox{number}}~]
    {s_1.\type=\jnum & s_2.\type=\jnum & r_1=\textit{schema2range}(s_1) & r_2=\textit{schema2range}(s_2)}
    {\delim{\{}\allof\punct{:}\delim{[}s_1,s_2\delim{]\}} -> 
     \textit{range2schema}(r_1 \cap r_2) \cup
     \delim{\{}\jmulof\punct{:}\textit{lcm}(s_1.\jmulof, s_2.\jmulof)\delim{\}}}
    \vspace{1em}
    \inference[\parbox{8mm}{intersect array}~]
    {s_1.\type=\jarray & s_2.\type=\jarray &
     s_1.\jitems=\delim{[}s_{1_1},\ldots,s_{1_k}\delim{]} & s_2.\jitems=\delim{[}s_{2_1},\ldots,s_{2_m}\delim{]}\\
     n=\textit{max}(k,m)}
    {\begin{array}{l@{}l@{}}\delim{\{}\allof\punct{:}\delim{[}s_1,s_2\delim{]\}} -> \delim{\{}
       & \type \punct{:} \jarray,\\
       & \jminitems \punct{:} \textit{max}(s_1.\jminitems, s_2.\jminitems),\\
       & \jmaxitems \punct{:} \textit{min}(s_1.\jmaxitems, s_2.\jmaxitems),\\
       & \begin{array}{@{}l@{}}
          \jitems \punct{:} \delim{[}
          \delim{\{}\allof\punct{:}\delim{[}s_{1_1} \parallel s_1.\jadditems, s_{2_1} \parallel s_2.\jadditems\delim{]\}},\\
          \hspace*{11mm}\ldots,\\
          \hspace*{11mm}\delim{\{}\allof\punct{:}\delim{[}s_{1_n}\!\parallel s_1.\jadditems, s_{2_n}\!\parallel s_2.\jadditems\delim{]\}]},\end{array}\\
       & \jadditems\punct{:}\delim{\{}\allof\punct{:}\delim{[}s_1.\jadditems,s_2.\jadditems\delim{]\}},\\
       & \juitems\punct{:}s_1.\juitems\wedge s_2.\juitems\delim{\}}
     \end{array}}
    \vspace{1em}
    \inference[\parbox{8mm}{intersect object}~]
    {s_1.\type=\jobj & s_2.\type=\jobj}
    {\begin{array}{l@{}l@{}}\delim{\{}\allof\punct{:}\delim{[}s_1,s_2\delim{]\}} -> \delim{\{}
       & \type \punct{:} \jobj,\\
       & \jminp \punct{:} \textit{max}(s_1.\jminp, s_2.\jminp),\\
       & \jmaxp \punct{:} \textit{min}(s_1.\jmaxp, s_2.\jmaxp),\\
       & \jreq  \punct{:} s_1.\jreq \cup s_2.\jreq,\\
       & \jpatp \punct{:} s_1.\jpatp \cup s_2.\jpatp\delim{\}}
     \end{array}}
    \vspace{1em}
    \inference[intersect anyOf~]
    {s_2=\delim{\{}\anyof\punct{:}\delim{[}s_{2_1},\ldots,s_{2_n}\delim{]\}}}
    {\delim{\{}\allof\punct{:}\delim{[}s_1,s_2\delim{]\}} -> \delim{\{}\anyof\punct{:}\delim{[}\delim{\{}\allof\punct{:}\delim{[}s_1,s_{2_1}\delim{]\}},\ldots,\delim{\{}\allof\punct{:}\delim{[}s_1,s_{2_n}\delim{]\}]\}}}
    \caption{Simplification rules to eliminate \texttt{allOf} except for connective \texttt{not}.}
    \label{rules:elim_allof}
\end{figure*}

Figure~\ref{rules:elim_allof} shows simplification rules for $\allof$.
For example, the intersection type in Figure~\ref{fig:str4} yields the
simplified schema in Figure~\ref{fig:str5}.
Rule \emph{singleton allOf} rewrites a conjunct of just one schema
into that schema.
Rule \emph{fold allOf} turns an $n$-ary $\allof$ into a binary one, so
the remaining rules need to handle only the binary case.
Rule \emph{intersect heterogeneous types} returns the bottom type because
intersection of incompatible types is the empty set, so the remaining
rules only need to handle homogeneously-typed schemas.
Rule \emph{intersect null} rewrites two nulls to one null.
Rule \emph{intersect boolean} uses the intersection of enumerations.
Rule \emph{intersect string} uses the intersection of regular expressions.
Rule \emph{intersect number} uses helper functions
\textit{schema2range} and \textit{range2schema} to convert back and
forth between number schemas and mathematical ranges, and \textit{lcm}
to compute the least common multiple of the $\jmulof$ constraints,
where \textit{lcm} handles undefined arguments by returning the other
argument if defined, or an undefined value if both arguments are
undefined.
Rule \emph{intersect array} takes advantage of the canonical form,
where $\jitems$ are always specified as lists, to compute an item-wise
intersection; undefined per-item schemas default to $\jadditems$.
Rule \emph{intersect object} simply picks up the union of the $\jpatp$
keywords, relying on the rule for objects with overlapping $\jpatp$ to
make them disjoint again later.
Finally, rule \emph{intersect anyOf} pushes conjunctions through
disjunctions by using the distributivity of intersection over union.
We choose not to push intersections through negations because we
prefer the end result of simplification to resemble distributive
normal form to the extent possible.

\begin{figure*}\small
    ~
    \inference[singleton anyOf~]
    {s.\anyof = \delim{[}s_1\delim{]}}
    {s -> s_1}
    \hspace{1em}
    \inference[fold anyOf~]
    {s.\anyof = \delim{[}s_1,s_2,\ldots,s_n\delim{]} & n \geq 3}
    {s -> \delim{\{}\anyof\punct{:}\delim{[}s_1, \delim{\{}\anyof\punct{:}\delim{[}s_2,\ldots,s_n\delim{]\}]\}}}
    \vspace{1em}
    \inference[union null~]
    {s_1.\type=\jnull & s_2.\type=\jnull}
    {\delim{\{}\anyof\punct{:}\delim{[}s_1,s_2\delim{]\}} -> \delim{\{}\type\punct{:}\jnull\delim{\}}}
    \vspace{1em}
    \inference[union boolean~]
    {s_1.\type=\jbool & s_2.\type=\jbool}
    {\delim{\{}\anyof\punct{:}\delim{[}s_1,s_2\delim{]\}} -> \delim{\{}\type\punct{:}\jbool, \enum\punct{:}s_1.\enum\cup s_2.\enum\delim{\}}}
    \vspace{1em}
    \inference[union string~]
    {s_1.\type=\jstr & s_2.\type=\jstr}
    {\delim{\{}\anyof\punct{:}\delim{[}s_1,s_2\delim{]\}} -> \delim{\{}\type\punct{:}\jstr, \jpattern\punct{:}s_1.\jpattern\cup s_2.\jpattern\delim{\}}}
    \vspace{1em}
    \inference[union number~]
    {s_1.\type=\jnum & s_2.\type=\jnum
     & r_1=\textit{schema2range}(s_1) & r_2=\textit{schema2range}(s_2)\\
     r_1\cap r_2 \neq \emptyset}
    {\begin{array}{@{}l@{}l@{}}
      \delim{\{}\anyof\punct{:} &
        \delim{[}s_1,s_2\delim{]\}} -> \delim{\{}\anyof\punct{:}\delim{[}\\
      & \textit{range2schema}(r_1 \cap r_2) \cup \delim{\{}\jmulof\punct{:}\textit{gcd}(s_1.\jmulof, s_2.\jmulof)\delim{\}},\\
      & \textit{range2schema}(r_1 \setminus r_2) \cup \delim{\{}\jmulof\punct{:}s_1.\jmulof\delim{\}},\\
      & \textit{range2schema}(r_2 \setminus r_1) \cup \delim{\{}\jmulof\punct{:}s_2.\jmulof\delim{\}]\}}
    \end{array}}
    \caption{Eliminating overlapping \texttt{anyOf}, except for connectives \texttt{not}
      and \texttt{allOf}, and types \texttt{array} and \texttt{object}.}
    \label{rules:elim_anyof}
\end{figure*}

Figure~\ref{rules:elim_anyof} shows simplification rules for $\anyof$.
In contrast to intersection, union allows incompatible types, e.g.,
\texttt{string} or \texttt{null} as in Figure~\ref{fig:str5}. Fortunately, such
heterogeneous unions are non-overlapping for inhabited schemas, so
they can be handled later during subschema checking.
Rule \emph{singleton anyOf} rewrites a disjunct of just one schema
into that schema.
Rule \emph{fold anyOf} turns an $n$-ary $\anyof$ into a binary one so
the remaining rules only need to handle the binary case.
Rule \emph{union null} rewrites two nulls to one.
Rule \emph{union boolean} uses the union of enumerations.
Rule \emph{union string} uses the union of regular expressions.
Rule \emph{union number} turns a binary union with overlap into a
ternary non-overlapping union. In other words, while it does not
eliminate the union of number schemas, it does simplify subschema
checking by at least making the union disjoint so it can be checked
elementwise.
Unfortunately, JSON Schema is not closed under union for types number,
array, and object. For example, the union of
\lstinline|{type:number,minimum:0}| and
\lstinline|{type:number,multipleOf:1}| is
$\mathbb{R}^{+}\cup\mathbb{Z}$, which cannot be expressed
in JSON schema without $\anyof$. There are similar counter-examples
for arrays and objects.  The case of unioned number schemas is handled
later in subschema checking.
As mentioned earlier, we would like simplification to end in schemas
that resemble a distributive normal form, so we choose not to push
$\anyof$ through $\allof$ or $\jnot$.

\subsection{JSON Subschema Checking}\label{sec:subtyping}

Given two canonicalized and simplified schemas, the third step of our approach checks whether one schema is a subtype of the other.
Figure~\ref{fig:subschema} presents inference rules defining the subschema
relation on canonical, simplified schemas. All rules are algorithmically
checkable, and all rules except for \emph{schema uninhabited} are
type-directed.  To simplify their presentation, some of the rules use
quantifiers, but all quantifiers are bounded and can thus be checked
via loops.

\begin{figure*}\small
~
\inference[subschema uninhabited~]
{\neg \inhabited(s_1)}
{s_1 <: s_2}\\
\vspace*{1em}~
\inference[subschema non-overlapping anyOf~]
{\forall i\in\{1..n\}, \exists j\in\{1..m\},  s_i <: t_j
& \textit{nonOverlapping}(\delim{[}t_1, \dots, t_m\delim{]})}
{\delim{\{}\anyof \punct{:} \delim{[}s_1, \dots, s_n\delim{]\}} <: \delim{\{}\anyof \punct{:} \delim{[}t_1, \dots, t_m\delim{]\}}}
\vspace{1em}
\inference[subschema null~]
{s_1.\type = \jnull & s_2.\type = \jnull}
{s_1 <: s_2}
\vspace{1em}
\inference[subschema boolean~]
{s_1.\type = \jbool & s_2.\type = \jbool &
s_1.\enum \subseteq s_2.\enum}
{s_1 <: s_2}
\vspace{1em}
\inference[subschema string~]
{s_1.\type = \jstr & s_2.\type = \jstr &
 s_1.\jpattern \subseteq s_2.\jpattern}
{s_1 <: s_2}
\vspace{1em}
\inference[subschema number~]
{ \forall i\in\{1..k\}, \jnot\notin\dom{s_i} \wedge s_i.\type=\jnum\\
  \forall i\in\{k+1..n\}, \jnot\in\dom{s_i} \wedge s_i.\jnot.\type=\jnum\\
  \forall i\in\{1..l\}, \jnot\notin\dom{t_i} \wedge t_i.\type=\jnum\\
  \forall i\in\{l+1..n\}, \jnot\in\dom{t_i} \wedge t_i.\jnot.\type=\jnum\\
  \subrange(([s_1,\ldots,s_k], [s_{k+1},\ldots,s_n]), ([t_1,\ldots,t_l], [t_{l+1},\ldots,t_m]))
}
{\delim{\{}\allof \punct{:} \delim{[}s_1,\ldots,s_k,s_{k+1},\ldots,s_{n}\delim{]\}} <: \delim{\{}\allof \punct{:} \delim{[}t_1,\ldots,t_l,t_{l+1},\ldots,t_m\delim{]\}}}
\vspace{1em}
\inference[subschema array~]
{s_1.\type=\jarray & s_2.\type=\jarray\\
s_1.\jminitems >= s_2.\jminitems & s_1.\jmaxitems <= s_2.\jmaxitems \\
s_1.\jitems=\delim{[}s_{1_1},\ldots,s_{1_k}\delim{]} & s_2.\jitems=\delim{[}s_{2_1},\ldots,s_{2_m}\delim{]}\\
\forall i\in[0,\ldots,\textit{max}(k,m)+1],\;
s_{1_i} \parallel s_1.\jadditems <: s_{2_i} \parallel s_2.\jadditems\\
s_2.\juitems \implies \left(s_1.\juitems \lor \alldisjointitems(s_1)\right)}
{s_1 <: s_2}
\vspace{1em}
\inference[\parbox{10mm}{subschema object}~]
{s_1.\type=\jobj & s_2.\type=\jobj\\
s_1.\jminp >= s_2.\jminp & s_1.\jmaxp <= s_2.\jmaxp \\
s_1.\jreq \supseteq s_2.\jreq\\
\forall p_1\!:\!s_{p_1} \in s_1.\jpatp, p_2\!:\!s_{p_2} \in s_2.\jpatp,
p_1\cap p_2\neq\emptyset \implies s_{p_1} <: s_{p_2}}
{s_1 <: s_2}
\caption{JSON Schema subtype inference rules.\label{fig:subschema}}
\end{figure*}

Rule \emph{schema uninhabited} states that an uninhabited schema is a
subtype of any other schema.
It uses an auxiliary $\inhabited$ predicate, which is elided for space
but easily
computable for primitives (recall that emptiness is decidable for
regular languages).  For structures, the predicate ensures that the schemas of all
required components are inhabited. For logic connectives, the
predicate is more involved but decidable.
The rule for uninhabited types is the only rule that is not type-directed.
Because canonicalization generally separates schemas by type, all other rules check same-typed schemas.
We can handle uninhabited schemas independently of their type because there is no actual data of that type that would require type-specific reasoning.
%For example, an inhabited string schema clearly cannot be a subschema of a number schema, since strings are not numbers.

Rule \emph{subschema non-overlapping anyOf} handles \anyof{} schemas
for the cases where simplification eliminates overlapping unions.
Helper function \textit{nonOverlapping} checks for unions of arrays
and objects and conservatively assumes that those might overlap.
In the non-overlapping case, it suffices to check the component schemas
independently.  For each schema on the left, we require a
same-typed super schema on the right.

%Rules \emph{subschema null}, \emph{subschema boolean}, and \emph{subschema string} are simple.

Rule \emph{subschema number} is the most complicated of the subtype rules
for primitive types due to \jmulof{} constraints. The simplifier
cannot push
negation through \jmulof{} constraints, and it cannot combine \allof{}
combinations of such negated schemas.  As a result, the rule
has to handle multiple
such constraints on both sides of the relation,
with or without negation.  We treat simple
number schemas as single-element \allof{}s for consistency.  This rule
verifies that any number allowed by the set of constraints on the left
is also allowed by the set of constraints on the right using an
auxiliary $\subrange$ relation, which is sketched in the following.

% \ahx{The subrange in the rule is very confusing. 
%     Also having the allOf on the bottom part of the rule is very unclear and 
%     not inline with the rest of the rules?
%     May be we revert back to the simpler rule we had earlier, and make the negated mulOf discussion in the text only?}
The $\subrange$ relation first normalizes all schema range bounds by
rounding them to the nearest included number that satisfies its
\jmulof{} constraint.  For each side, it then finds the least and
greatest finite bound used.  Every unbounded schema is split into two
(or three for totally unbounded) schemas: one (or two) that are
unbounded on one side, with the least/greatest bound as the other
bound.  The ``middle'' part is bounded.  All these derived schemas
keep the original \jmulof.  The bounded schemas can all be checked
(exhaustively if needed).  For the unbounded schemas, we can
separately check the positive and negative schemas, since they do not
interact in interesting ways over unbounded sets.  If $PL$ and $PR$
are the left and right positive schemas, and $NL$ and $NR$ are the
left and right negative schemas, we verify that the constraints
divide each other:
\begin{align*}
\forall_{\textit{pl}\in\textit{PL}}, \exists_{\textit{pr}\in\textit{PR}},\; \textit{pl}.\jmulof \textrm{ mod } \textit{pr}.\jmulof = 0\\
\forall_{\textit{nr}\in\textit{NR}}, \exists_{\textit{nl}\in\textit{NL}},\; \textit{nr}.\jmulof \textrm{ mod } \textit{nl}.\jmulof = 0
\end{align*}

Rule \emph{subschema array} checks two array schemas.
%The minimum and maximum
%items of the former need to be not less than (resp.\ not greater than)
%the latter.  
The left array size bounds should be within the size bounds of the right array.
Additionally, the schema of every item specified in the
former needs to be a subschema of the corresponding specification in
the latter.  If a schema is not explicitly provided, the schema
provided by \jadditems{} is used. Recall that canonicalization adds in
a default \jadditems{} schema if it was not specified.
Additionally, if the right side specifies that the items must be
unique, then the left needs to either specify the same or implicitly
enforce this. For example,
\lstinline|{type:array,items:[{enum:[0]},{enum:[1]}]}| is a subschema
of \lstinline|{type:array,uniqueItems:true]}|.
The $\alldisjointitems$ predicate checks for this by
first obtaining the set of all the effective item schemas: every item
schema for an index within the specified min/max bounds, and
$\jadditems$ if any allowed indices are unspecified.  It then verifies
that the conjunction of all pairs of effective items schemas are
uninhabited.

Rule \emph{subschema object} checks two object schemas. It first
verifies that the number of properties of both sides have the appropriate
relation, and that the left side requires all the keys that the right
side requires. Next, for every regular expression pattern $p_1$ on the
left, if there is an overlapping regular expression pattern $p_2$ on the
right, it checks that the corresponding schemas are subschemas.
This check can be done separately for one pattern at a time thanks to
the fact that canonicalization eliminates overlapping pattern
properties.

%\mpx{If one schema is found to not be a subschema of another, what is shown to a user to explain this result?}

%%% Local Variables:
%%% mode: latex
%%% TeX-master: "../main"
%%% End:

\section{Implementation}\label{sec:impl}

We implemented our subschema checker as a Python tool in around 2,000
lines of code.  The implementation builds upon the \textsf{jsonschema}
library\footnote{\url{https://github.com/Julian/jsonschema}} to
validate schemas before running our subtype checking, the
\textsf{greenery}
library\footnote{\url{https://github.com/qntm/greenery}} for computing
intersections of regular expressions, and the \textsf{jsonref}
library\footnote{\url{https://github.com/gazpachoking/jsonref}} for
resolving JSON schemas references.

%\subsection{Limitations}

\section{Evaluation}\label{sec:evaluation}

This section evaluates the implementation of our JSON subschema
checker, which we refer to as \subschema.
It answers the following research questions:

\begin{description}
  \item[RQ$_1$] How correct is \subschema{} in practice?
  \item[RQ$_2$] How does \subschema{} fare against existing work?
  \item[RQ$_3$] How complete is \subschema{} in practice?
  \item[RQ$_4$] How efficient is \subschema?
\end{description}

\subsection{Experimental Setup}

We evaluate our subschema checker on four datasets of JSON schemas from different domains: JSON schema official test suite, \wpt{}, \kub{}, and \lale.

The official test suite for JSON Schema \mbox{draft-04}~\footnote{https://github.com/json-schema-org/JSON-Schema-Test-Suite} is a widely-used test suite for JSON Schema validators and provides 146 schemas and 531 tests that offer full coverage of the JSON Schema specification. 
Each test provides a JSON document $d$ to be validated against a specific schema $s$
and the expected correct validation behavior $\textit{res} \in \{True, False\}$, i.e., tests are of the form 
$\valid(d,s) = \textit{res}$.
These schemas are designed to cover the JSON schema features, but they are not representative for real-world schemas.
Moreover, these schemas were not designed to test the subschema relation we define in this work, so we 
only use this test suite to evaluate the canonicalization and simplification steps (Sections~\ref{sec:canonicalization} and~\ref{sec:simplification}) of our approach.

The three other datasets, \wpt{}, \kub{}, and \lale, are used for evaluating the full \subschema{}
approach. \wpt{} is a collection of schemas describing content used by 
the Washington Post within the \emph{Arc Publishing} content creation and management system.\footnote{\url{https://github.com/washingtonpost/ans-schema}}
\kub{} is the set of JSON schemas describing the OpenAPI specifications 
for \emph{Kubernetes}, an open-source system for automating deployment, 
scaling, and management of containerized applications.\footnote{\url{https://kubernetes.io/}}
Since OpenAPI specifications contain more information beyond the schemas for the REST API endpoints, 
we use a set of JSON schemas extracted from them.\footnote{\url{https://kubernetesjsonschema.dev/}}
Specifically, we used the standalone flavor of schemas where \jref{} have been resolved to local files.
The \lale{} dataset is a set of schema pairs from the \emph{Lale}
open-source project\footnote{\url{https://github.com/ibm/lale}}, which
is a Python library for type-driven automated machine learning.

The second and third datasets, \wpt{} and \kub{}, comprise several versions.
We apply our subschema checker across each pair of consecutive
versions of the same schema that introduces some textual modification,
to spot whether a change may impact the compatibility of the
corresponding systems.
% That is, for two schema versions $s_{v_n} <: s_{v_m}$ or $s_{v_n} :> s_{v_m}$ 
% where $n < m$. 
%
We use the third dataset, \lale{}, to find type errors in AI pipelines
where wrong operators could be applied to specific datasets. We
consider 4 \lale{} operators and 7 datasets, yielding 28 schema pairs.

Table~\ref{table:dataset} shows statistics for the three datasets that we use for the full evaluation of \subschema{}.
For example, \kub{} has in total 124 versions, with a total of 86,461 schemas.
Due to the additions and deletions of schemas between pairs of consecutive versions, 
the total number of pairs of schemas across all pairs of subsequent versions is 82,814.
Finally, since not every new version of an API modifies every schema, 
we only keep pairs of non-equal files.
Overall, the total number of pairs of schemas is 8,548.
Many of the schemas are of non-trivial size, with an average of 56KB and a maximum of 1,047KB.
The first dataset, the JSON schema test suite, is omitted from Table~\ref{table:dataset} since it does not have pairs of versions and we only use it to evaluate the canonicalization and simplification steps.

%MIN file size: 94 B
%Max file size: 1047.4203454894434 KB
%AVG file size: 57724.30991784038 B
%AVG file size: 56.371396404140995 KB
\begin{table}
    \caption{Dataset details.}
    \vspace*{-2mm}
    \label{table:dataset}
    \setlength{\tabcolsep}{5pt}
    \small
    \begin{tabular}{@{}lrrrr@{}}
        \toprule
%        & & & & & \multicolumn{2}{c}{\% of failures} \\
        \textbf{Dataset} & \textbf{Versions} & \textbf{Schemas} & \textbf{Schema pairs} & \textbf{Schema pairs used for evaluation} \\
%        & \textbf{Subschema} & \textbf{isSubset}\\
        \midrule
        \wpt{} & 28 & 2,604 & 2,411 & 2,060 \\ % & 23.16\%& 100\%\\
        \kub{} & 124 & 86,461 & 82,814 & 6,460 \\ % & 0.35\% & 0\% \\
        \lale{} & -- & -- & 28 & 28 \\% & 0\% & 100\%\\
        \midrule
        \textbf{Total} & & & & 8,548 \\
        %\midrule
        %\multicolumn{2}{l}{\textbf{File size}} & \textbf{Min.} & \textbf{Max.} & %\textbf{Average.} \\
        %\midrule
        %& & 94B & 1,047KB & 56KB \\
        \bottomrule
    \end{tabular}
%    \vspace{-.5em}
\end{table}

All experiments are performed on an Intel Core i7-4600U CPU (2.10GHz) machine with 16GB
of memory running Ubuntu 18.04 (64-bit).

\subsection{Correctness in Practice}

For RQ$_1$, we evaluate the correctness of the two main steps of our approach: 
canonicalization and simplification (canonicalization), and subtype checking.

\subsubsection{Canonicalization}
Canonicalization and simplification aim at producing a valid and simpler canonical schema that
is semantically equivalent to the input schema.
To test validity, \subschema{} checks the canonicalized schema against
the meta-schema of JSON Schema using an off-the-shelf JSON schema
validator (Section~\ref{sec:impl}).
Across our entire dataset, there is no single case where this check fails.
%In Section~\ref{}, we discuss the cases where \subschema{} fails to produce a result, 
%which does not involve failing this sanity check on hundreds of schemas.

To check that canonicalization and simplification preserve the semantics of the original input 
schema, we apply the canonicalization step to all schemas in the JSON Schema official test suite.
We tested whether:
\mbox{$\forall s, \forall d, \valid(d,s) = \textit{res} \implies \valid(d,\canonical(s)) = \textit{res}$}.
for schema $s$, JSON document $d$, and outcome $res$.
%
% Resuls were not accurate, updating them
%Our canonicalizer successfully canonicalizes 129 out of the 146 schemas.
In all cases except one where \subschema{} yields a canonicalized schema, this new schema passes all relevant tests in the JSON schema test suite.
This single case, we believe, is due to the ambiguity of the specification of JSON schema and hence, 
a mismatch between our own interpretation and the interpretation of the JSON schema validator 
of the semantics of the \lstinline|allOf| connector when combined with the \lstinline|additionalProperties| object constraint.

These results, of course, show only that canonicalization in most of the cases does not yield an invalid or a more strict schema than the input schema.
The following experiments on the correctness of subtype checking rely on correct canonicalization, and hence, provide additional evidence.

\subsubsection{Subtype Checking}{\label{sub:correctness:subtyping}}

\paragraph{Self-equivalence.}
As an automated, large-scale correctness check of the subtype checking, we perform a simple sanity check that asks \subschema{} whether a schema is equivalent to itself (Definition~\ref{def:equiv}).
We randomly sample 1000 schemas from the \kub{} dataset and run our subschema checker using the same schema on both sides of the subtype relation, 
i.e., checking for a schema $s$ whether $s <: s$.
Our subtype checking does not rely on any sort of structural equality.
Therefore, in this setup, our implementation is oblivious to the fact that both schemas are the same, so it 
canonicalizes both schemas and then performs subtype checking.
In all 1,000 samples, this test passes correctly.

\paragraph{Comparison against a ground truth.}
For further validation, we compare the results of \subschema{} against a ground truth.
Specifically, we gather pairs of schemas, along with their expected subtype relationship, in three ways.
First, we randomly sample pairs that are textually different and manually assess their subtype relationship.
Second, we sample consecutive versions of schemas from \wpt{} and \kub{} and manually assess their subtype relationship.
For example, we consider versions 0.7.0 and 0.7.1 of the \texttt{utils/named\_entity} schema from the \wpt{} dataset.
This sample of pairs represents the usage scenario where our approach checks whether an evolving API specified through a JSON schema may break an application.
Third, for the 28 schema pairs from \lale{}, the ground truth is
whether or not the corresponding machine-learning operator throws an
exception when training on the corresponding dataset.
By checking the schemas statically, our subtype checker can avoid such runtime errors.
To focus on the correctness of the subtype checker output, we sampled pairs of schemas from 
the set of schemas where our approach yields a decision, i.e., ignoring the few cases where \subschema{} cannot decide on the subschema relation.

% For lale, expected output is TP=12; TN=16
% issbuset observed output: FP=0; FN=8; TP=3; TN=10; failures=7
\begin{table}
	\caption{Effectiveness of \subschema{} and comparison to the existing \issubset{} tool.}
    \vspace*{-2mm}
	\label{table:correctness}
	\small
	\setlength{\tabcolsep}{7pt}
	\centerline{\begin{tabular}{@{}lr rrrrr | rrrrr@{}}
		\toprule                                                                            
		& & \multicolumn{5}{c|}{\subschema} & \multicolumn{5}{c}{\issubset}\\
		\cmidrule{3-12}
		& \textbf{Pairs} & \textbf{Fail} & \textbf{TP} & \textbf{TN} & \textbf{FP} & \textbf{FN} &
		\textbf{Fail} & \textbf{TP} & \textbf{TN} & \textbf{FP} & \textbf{FN}\\
		\midrule
		$<:$           &  35 & 0 &  29 &  6 & 0 & 0 & 10 &   9 &  0 &  6 & 10\\
		$:>$           &  35 & 0 &  31 &  4 & 0 & 0 & 10 &  21 &  0 &  4 &  0\\
		$\equiv$       & 100 & 0 & 100 &  0 & 0 & 0 & 50 &  27 &  0 &  0 & 23\\
		$\not\equiv$   & 100 & 0 &  63 & 37 & 0 & 0 &  0 &  63 &  0 & 37 &  0\\
		\midrule
		\lale{}        &  28 & 0 &  12 & 16 & 0 & 0 &  7 &   3 & 10 &  0 &  8\\
		\midrule
		\textbf{Total} & 298 & 0 & 235 & 63 & 0 & 0 & 77 & 123 & 10 & 47 & 41\\
		\bottomrule
	\end{tabular}}
%    \vspace{-.5em}
\end{table}

In total, we gathered 298 pairs with a ground truth subtype
relationship, as summarized in Table~\ref{table:correctness}.  The
$<:$, $:>$, $\equiv$, and $\not\equiv$ symbols represent what test we
performed on each pair. For example, for each pair
\mbox{$\langle s,t\rangle$} in the $<:$ row, the ground truth
indicates whether $s<:t$ holds (positive,~P) or not (negative,~N).
The \subschema{} part of the table shows the results of applying our
subschema checker to each pair.  The TP, TN, FP, and FN columns
indicate the true positives, true negatives, false positives, and
false negatives, respectively.  For example, TN means that the tool
produces the correct result (T for true) and that the ground truth
indicates that the relationship being tested is not expected to hold
(N for negative).  Our tool produces
the correct results for all 298 schema pairs in the ground truth.

%\mpx{How come with we have zero failures here, even though the tool is incomplete as we acknowledge in Section~\ref{sub:completeness}?}\ahx{Incompleteness hits rare corner cases, which did not happen in this sample.}

\subsection{Comparison to Existing Work}
\label{sec:evalComparison}

As we discuss in Section~\ref{sec:related}, our work is the first to 
%\mpx{Not ``formally'' enough for some reviewers -- remove the claim?}
define the subschema relation on JSON schemas and present an algorithm to perform this check for a large subset of JSON schema features.
Therefore, to our knowledge, there is no academic work that we can compare against.
However, since our work is motivated by the practical need for subtyping JSON schemas,
for RQ$_2$,
we compare our implementation to the closest developer tool we could find, 
\textsf{is-json-schema-subset} (\issubset)~\cite{haggholm_2019}.
The \issubset{} tool is written in TypeScript and its documentation states the same goal as ours: 
``Given a schema defining the output of some process A, and a second schema defining 
the input of some process B, will the output from A be valid input for process B?''
We use the most recent version, which is 1.0.6.

We first run \issubset{} on the three datasets, \wpt, \kub, and \lale.
It fails to run on some schemas from \wpt{} and \lale{}
due to unsupported schema versions, although the \issubset{} documentation
does not describe such a limitation.
Next, we compare the correctness results of \issubset{} against our \subschema{}.
The right part of Table~\ref{table:correctness} shows the correctness 
results for \issubset{} using the methodology described in Section~\ref{sub:correctness:subtyping}.

The first observation is that \issubset{} produces a non-negligible number of true
positives, which means it indeed captures some of the semantic of the subtyping relation.
However, \issubset{} also produces 47 false positives and 41 false negatives, i.e., gives a wrong answer to a subtype query.
Overall, the existing tool gives a wrong answer in 40\% of the cases where the tool does not fail.

To get a better understanding of the low recall, we inspected the code of \issubset{}
and tested it on simple hand-crafted schemas. 
We find that although the tool performs some simple semantic checks, e.g.,
it correctly reports \lstinline|{'type': 'integer'}| $<:$ \lstinline|{'type': 'number'}|, 
it lacks the ability to capture the richness of JSON schema in many ways.
For instance, it fails to detect \lstinline|{'type': ['string','null']}| $\equiv$ \lstinline|{'type': ['null', 'string']}|, and is oblivious to uninhabited schemas,
such as \lstinline|{'type': 'string', 'enum': [1]}|.

\subsection{Completeness in Practice}\label{sub:completeness}
Being complete in practice is difficult. 
To balance completeness and effort, there is a set of features our approach
currently cannot deal with.
%\mpx{How does his relate to the theoretical incompleteness according to Section~\ref{sec:algo}?}
As discussed in Section~\ref{sec:algo}, our approach cannot reason about negation and complement of object, array, and numeric with \lstinline|multipleOf| schemas.
Therefore, for RQ$_3$, we report on two dimensions.
The first is the pervasiveness of different validation keywords and which of them \subschema{} supports.
The second dimension is failure cases of \subschema, i.e, cases where we do not produce a subtype decision, due to a limitation of the approach.

\subsubsection{Pervasiveness of Validation Keywords and Supported Features}
\begin{figure}
  \centering
  \includegraphics[page=1,width=\linewidth]{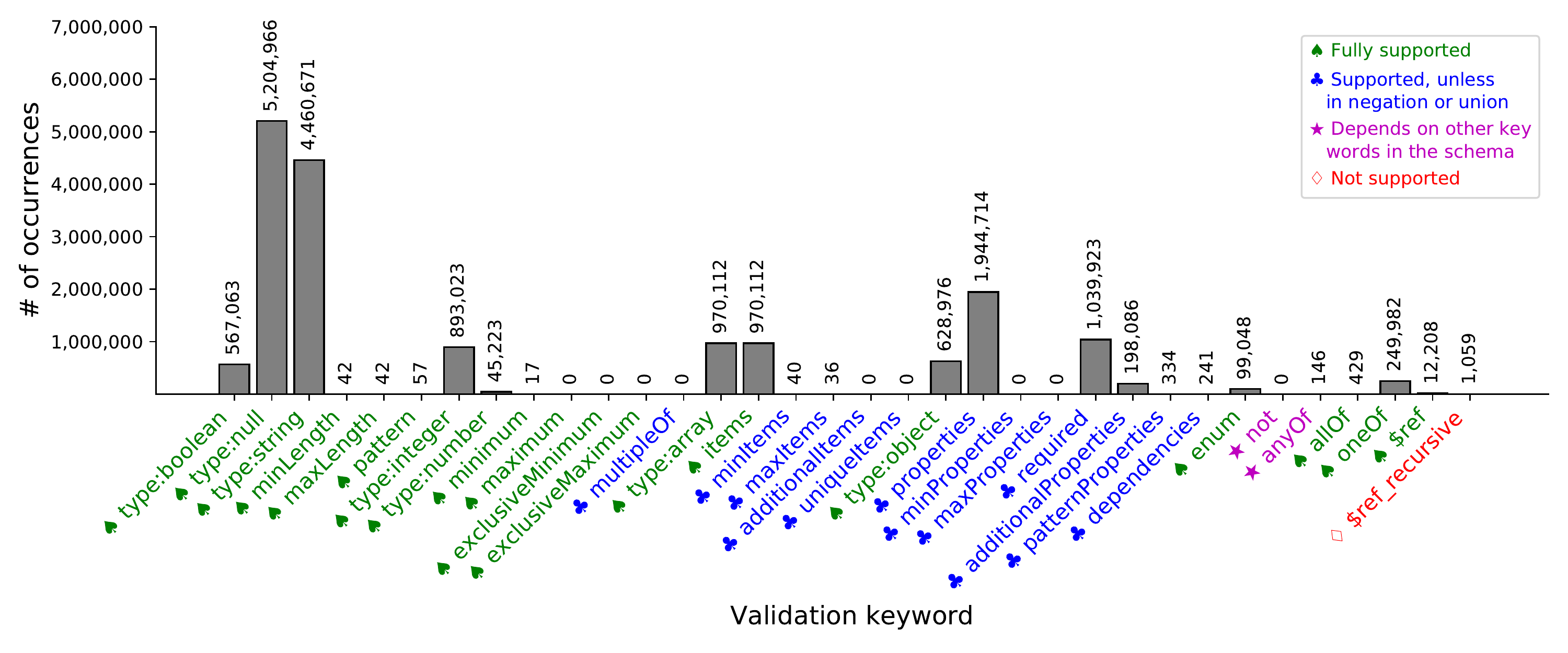}
  \caption{Prevalence of JSON schema validation keywords in practice and supported features in \subschema{}.}
  \label{fig:prevalence}
\end{figure}

Figure~\ref{fig:prevalence} shows the frequency of validation keywords across 
all schemas in the \kub{} and \wpt{} datasets.
Validation keywords on the x-axis are sorted by their relevance to each schema type and  according to the order of keywords in Table~\ref{table:jschema}.
The figure shows that \subschema{} indeed supports the majority 
of JSON schema features which are used in practice

We observe that JSON schema types \lstinline|null| and \lstinline|string|
are the two most prevalent schema types present in the dataset.
Both types are fully supported in the subtype checking performed by \subschema{}
as indicated by the color code in Figure~\ref{fig:prevalence}.

The keywords \lstinline|properties|, \lstinline|additionalProperties|, and \lstinline|required| for specifying constrains on a JSON \lstinline|object| show up next
on the order of the number of use cases.
Of these keywords, \subschema{} fully supports \lstinline|properties|. The \lstinline|additionalProperties| and \lstinline|required| keywords are supported whenever they are not used in union schemas or negated schemas.
In general, disjunction of schemas happens rarely (146 occurrence among millions of occurrences of other keywords); while negated schemas are not used at all in our dataset of real-world schemas.

Worth noting here is that the counts in Figure~\ref{fig:prevalence} are for schemas that do not use the negated schema keyword \lstinline|not| at all, which is also evident from its frequency being 0.
The reason is that in the dataset of schemas, there is no single use of a negated schema.
In fact, the use of negation in JSON schemas is indeed highly discouraged since the purpose of schema validation is to constraint what is allowed rather than filtering out what is disallowed.
The NSA security guidelines for using JSON schemas also advises against using negated schemas for the same reason~\footnote{\url{https://apps.nsa.gov/iaarchive/library/reports/security_guidance_for_json.cfm}}.
That said, \subschema{} still supports the negation of all $\jprimitive{}$ except for the 
union and negation of numeric schemas with a \lstinline|multipleOf| constraint.
Overall, this shows that the incompleteness of \subschema{} rarely affects its usability in real world, on a large dataset of real-world schemas.

The only feature that is not supported at the moment is recursive references 
in schemas using \lstinline|$ref|.
Although our approach currently does not handle recursive schemas, we know theoretically 
that subtyping recursive types is decidable~\cite{amadio_cardelli_1993}.

\subsubsection{Unsupported Features in Practice}

The \subschema{} tool reports an error and terminates without yielding a decision when
an unsupported JSON schema feature is encountered.

Table~\ref{table:failures} shows the cases when \subschema{} fails on our dataset.
In total, out of 8,548 schemas pairs, the subschema checks fails for 5.69\% of the pairs.
The table shows three kinds of failures that happen in practice due to limitations of our approach.
The first and most dominant failure reason are circular and recursive schemas.
As indicated above, this is not an inherent limitation of \subschema{}.
The second case is the rare use case of negating objects schemas.
%We opted not to implement this negation of arrays and objects believing that they are rarely used.
%\ahx{To Avi/Martin. Can we say something like we showed this case to be not computable using the default set of schema validation keywords?}
As seen in Table~\ref{table:failures}, only 0.34\% of schema pairs fail due to the absence of this feature.
It is worth noting here is that the original schemas do not use negated schemas at all 
as seen in Figure~\ref{fig:prevalence}.
However, they were introduced as part of \subschema{} canonicalization phase (Section~\ref{sec:canonicalization}) where \lstinline|oneOf| constraints are re-written into disjunction of conjunctions with negations (Figure~\ref{rules:canonical}).
We have discussed in Section~\ref{sec:algo} that JSON schema is not closed under union or negation of object schemas.
The third case is when string schemas or object schemas
use non-regular language for specifying textual patterns using the keywords 
\lstinline|pattern| amd \lstinline|patternProperties|, respectively.
Inclusion in non-regular languages (e.g., regular expressions with positive and negative look-around) is undecidable and is beyond our scope.
%\mpx{Clarify what ``non-classical'' means.}
%\mpx{Does this mean that subschema checking for JSON schemas is undecidable in general?}

For eleven pairs of schemas, at least
one of the schemas had an invalid \jref{}. For two pairs of
schemas, at least one of the two files is not a valid JSON document.
These thirteen pairs are omitted from Table~\ref{table:failures}.
%\mpx{How about omitting these cases completely? If they are invalid or incorrect schemas, no need to use them in the evaluation.}

\begin{table}
\begin{minipage}{0.45\linewidth}
  \centering
  \caption{Reasons for incompleteness in practice.}
  \label{table:failures}
  \begin{tabular}{lrr}
    \toprule
    \textbf{Failure reason} & \textbf{Count} & \textbf{\%} \\
    \midrule
    %        \multicolumn{3}{c}{Unsupported feature} \\
    %        \midrule
    Recursive or circular \jref & 453 & 5.30\% \\
    Negated object schema       &  29 & 0.34\% \\
    Non-regular regex pattern   &   5 & 0.06\% \\
    % one can remove this part of the table
    \midrule
    %        \multicolumn{3}{c}{Other errors}\\
    %        \midrule
    %        Unresolved refs             &  11 & 0.13\% \\
    %        Invalid JSON file           &   2 & 0.02\% \\
    %        \midrule
    %        %
    %        \textbf{Total}              & 500 & 5.85\% \\
    %        \bottomrule
    \textbf{Total}              & 487 &5.69\% \\
    \bottomrule
  \end{tabular}
\end{minipage}
\hfill
\begin{minipage}{0.45\linewidth}
\centering
\includegraphics[scale=.35]{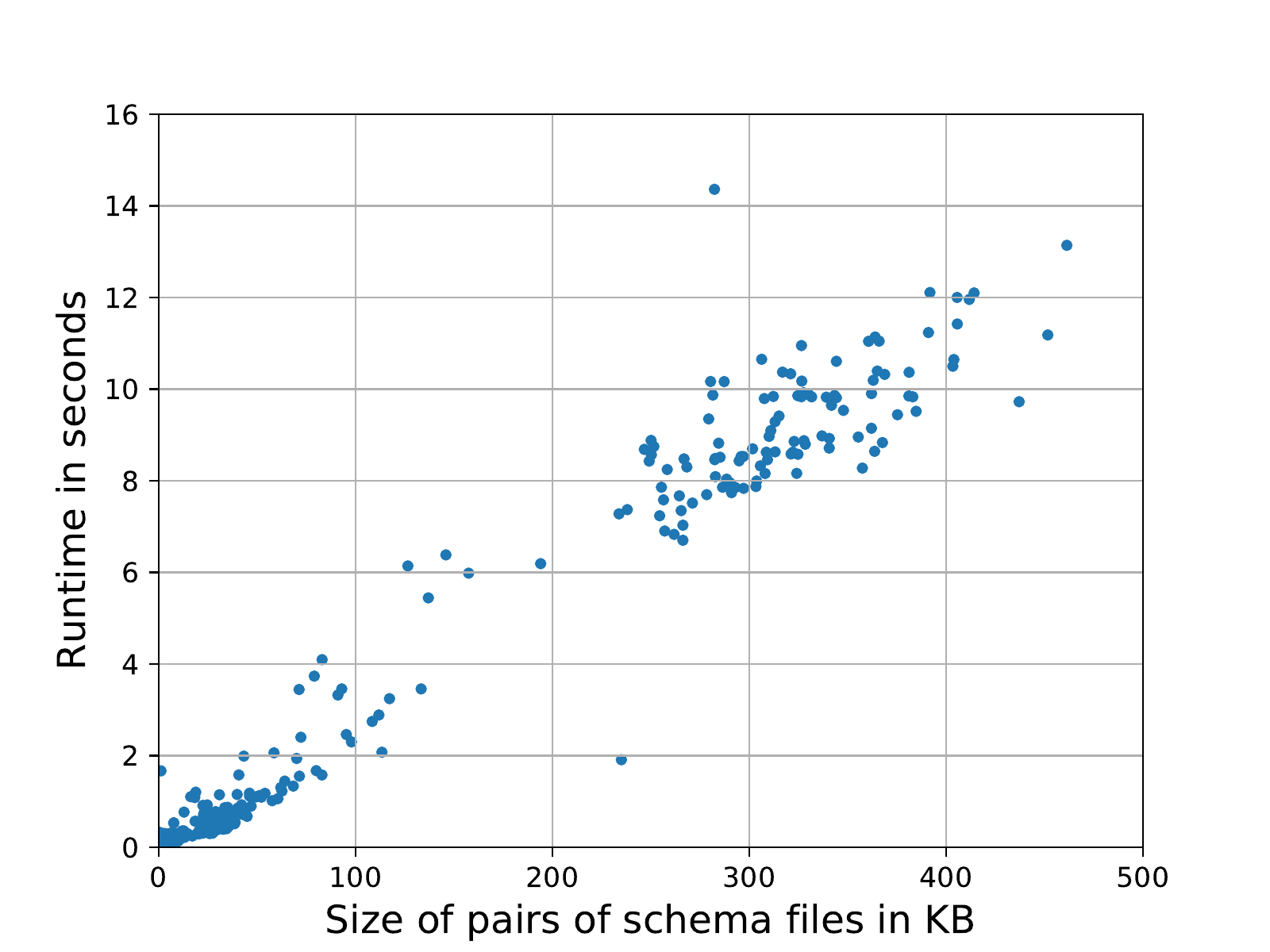}
\captionof{figure}{Efficiency of Subschema checking.}
\label{fig:sizeVstime}
\end{minipage}
\end{table}

%\begin{table}
%    \caption{Reasons for incompleteness in practice.}
%    \vspace*{-2mm}
%    \label{table:failures}
%    \small
%    \begin{tabular}{lrr}
%        \toprule
%        \textbf{Failure reason} & \textbf{Count} & \textbf{\%} \\
%        \midrule
%%        \multicolumn{3}{c}{Unsupported feature} \\
%%        \midrule
%        Recursive or circular \jref & 453 & 5.30\% \\
%        Negated object schema       &  29 & 0.34\% \\
%        Non-regular regex pattern   &   5 & 0.06\% \\
%        % one can remove this part of the table
%        \midrule
%%        \multicolumn{3}{c}{Other errors}\\
%%        \midrule
%%        Unresolved refs             &  11 & 0.13\% \\
%%        Invalid JSON file           &   2 & 0.02\% \\
%%        \midrule
%%        %
%%        \textbf{Total}              & 500 & 5.85\% \\
%%        \bottomrule
%        \textbf{Total}              & 487 &5.69\% \\
%        \bottomrule
%    \end{tabular}
%%    \vspace{-.5em}
%\end{table}

For the same incompleteness reasons described here, the canonicalizer failed to canonicalize 
22 schemas out of the 146  schemas in the JSON schema test suite.
However, the schemas in this test suite indeed cover the  entire language features of JSON 
schema but they do not represent the real-world use cases of JSON schema in practice.

\subsection{Efficiency}
To evaluate how fast our subschema checker is in practice for RQ$_4$,
we measure the time taken by subschema checks on a sample of 798 pairs 
of non-equal schemas from Table~\ref{table:dataset}.
We took every time measurement ten times and report the average.
Figure~\ref{fig:sizeVstime} shows the size of pairs of schema files in KB
against the time subschema checking takes in seconds.

In most cases, our subschema checker terminates within a few seconds
for moderately sized schemas, with time increasing
roughly linearly with the schema file size.
However, our subschema approach is lazy and terminates on the first violation 
of a subtyping rule.
On one pair of schemas in our dataset, eliminated from the figure for scaling sake,
it took around 2.8 minutes to terminate, which is not optimal for production.
%\mpx{Is this pair missing in the figure?}\ahx{yes}
We will explore how to improve on this, for instance, by on-demand canonicalization.

\section{Related Work}\label{sec:related}

\subsection{JSON Schema and Schema Subtyping}

Practitioners have significant interest in reasoning about the subtype relation of JSON schemas.
Section~\ref{sec:evalComparison} has an experimental comparison
against the strongest competitor among the available tools,
\textsf{is-json-schema-subset}~\cite{haggholm_2019}, which was developed
concurrently with our work.
Another closely related tool\footnote{\url{https://bitbucket.org/atlassian/json-schema-diff-validator}} relies on simple syntactic checks.
For example, that work considers a change as a breaking change whenever a node is removed from the schema.
As illustrated by Figure~\ref{fig:str}, removing nodes (and replacing them by others) may yield not only subtypes but even equivalent schemas.
Yet another existing tool\footnote{\url{https://github.com/mokkabonna/json-schema-compare}} checks whether two schemas are equivalent but does not address the subtyping problem.

\citet{pezoa_et_al_2016} formally define the syntax and semantics of JSON Schema, including the JSON validation problem.
An alternative formulation of JSON validation uses a logical formalism~\cite{DBLP:conf/pods/BourhisRSV17}.
\citet{DBLP:conf/dbpl/BaaziziCGS17} address the problem of inferring schemas for irregular JSON data, but their work does not use the JSON Schema standard we are targeting here.
None of the above pieces of work addresses the subschema problem.

There are other schema definition languages for JSON besides JSON Schema.
One popular alternative is the Swagger/OpenAPI specification language.\footnote{\url{https://swagger.io/}}
While similar to JSON Schema, it is not fully compatible.
The \textsf{swagger-diff}
tool\footnote{\url{https://github.com/civisanalytics/swagger-diff}}
aims at finding breaking API changes through a set of syntactic
checks, but does not provide the detailed checks that we do.
Avro is another schema definition language\footnote{\url{http://avro.apache.org/}}, which, however, does not specify a subschema relation.
We envision our work to help define subtype relations of these alternative schema definition languages.

% ignored for now; seems not very popular: https://github.com/inveniosoftware/doschema

% ignored for now; seems only vaguely related: https://github.com/mokkabonna/json-schema-merge-allof

\subsection{Type Systems for XML, TypeScript, and Python}

CDuce is a functional language designed for working with XML, which can reason about the types of XML documents and about subtype relations~\cite{DBLP:conf/icfp/BenzakenCF03}.
\cite{tozawa_hagiya_2003} address the problem of subschema checking for XML, where it is called schema containment. These approaches treat XML as tree automata, which is impossible for JSON, as JSON Schema is more expressive than tree automata~\cite{pezoa_et_al_2016}.

Both JavaScript and Python have a convenient built-in syntax for JSON
documents. Furthermore, there are type systems retrofitted onto both
JavaScript~\cite{bierman_abadi_torgersen_2014} and
Python~\cite{vitousek_et_al_2014}. Therefore, a reasonable question to
ask is whether JSON schema subtype queries could be decided by
expressing JSON documents in those languages and then using the
subtype checker of those type systems. Unfortunately, this is not the
case, since JSON Schema contains several features that those type
systems cannot express. For instance, JSON Schema supports negation,
\jmulof{} on numbers, and \jpattern{} on strings, none of which those
type systems support.

\subsection{Applications of Subschemas}
\label{sec:relatedWorkApplications}

One application of JSON subschema is for statically reasoning about breaking changes of web APIs.
A study of the evolution of such APIs shows that breaking changes are frequent~\cite{DBLP:conf/icws/LiXLZ13}.
Another study reports that breaking changes of web APIs cause distress among developers~\cite{DBLP:journals/jss/EspinhaZG15}.
Since JSON schemas and related specifications are widely used to specify data types, our subschema checker could help identify breaking changes already statically and on the schema-level, instead of relying on testing.

Data validation for industry-deployed machine learning pipelines is of crucial value as such pipelines are usually retrained frequently with new data. In order to validate incoming data, Google TFX~\cite{Baylor2017} synthesizes a custom data schema based on statistics from available data and uses this schema to validate future data instances fed to the TensorFlow pipeline~\cite{Breck2019}. 
Amazon production ML pipelines~\cite{Schelter2018} offer a declarative API which allow users to manually define desired constraints or properties of data. Then data quality metrics such as completeness and consistency are measured on real-time data with respect to the pre-defined constraints and anomalies are reported. 
Both systems are missing an explicit notion of schema subtyping.
For instance, TFX uses versioned schemas to track the evolution of inferred data schemas, and reports back to the user whether to update to a more (or less) permissive schema based on the historical and new data instances~\cite{Baylor2017}.
\textsc{Lale} uses JSON schemas to specify both correct ML pipelines and their search space of hyperparameters~\cite{hirzel_et_al_2019}.
The ML Bazaar also specifies ML primitives via JSON~\cite{smith_et_al_2019}.
Another type-based system for building ML pipelines is described by \cite{pilat_kren_neruda_2016}.
These systems could benefit from JSON subschema checking to avoid running and deploying incompatible ML pipelines.

%% \begin{alltt}\textcolor{red}{TODO}\scriptsize
%% 	- transmogrif.ai, https://www.slideshare.net/MatthewTovbin/meet-transmogrifai-open-source-automl-that-powers-einstein-predictions-136249954
%% \end{alltt}

%%% Local Variables:
%%% mode: latex
%%% TeX-master: "../main"
%%% End:

\section{Conclusion}\label{sec:conclusion}

This paper introduces a subtype checker for JSON Schema. There are
several features in JSON Schema that make subtype checking difficult,
including a full set of Boolean connectives, enumerations containing
values of possibly heterogeneous other types, regular expressions for
strings, and multiple-of constraints for numbers. Our checker is the
first to effectively handle these cases. The evaluation demonstrates that
the tool works well on a large set of examples of high real-world
importance, including web APIs, cloud computing, and artificial
intelligence.

\vfill

% Text of paper \ldots

% %% Acknowledgments
% \begin{acks}                            %% acks environment is optional
%                                         %% contents suppressed with 'anonymous'
%   %% Commands \grantsponsor{<sponsorID>}{<name>}{<url>} and
%   %% \grantnum[<url>]{<sponsorID>}{<number>} should be used to
%   %% acknowledge financial support and will be used by metadata
%   %% extraction tools.
%   This material is based upon work supported by the
%   \grantsponsor{GS100000001}{National Science
%     Foundation}{http://dx.doi.org/10.13039/100000001} under Grant
%   No.~\grantnum{GS100000001}{nnnnnnn} and Grant
%   No.~\grantnum{GS100000001}{mmmmmmm}.  Any opinions, findings, and
%   conclusions or recommendations expressed in this material are those
%   of the author and do not necessarily reflect the views of the
%   National Science Foundation.
% \end{acks}

%% Bibliography
\bibliography{bibfile}

%% Appendix
% \appendix
% \section{Appendix}

% Text of appendix \ldots

\end{document}